\documentclass[iop]{emulateapj}
\usepackage{graphicx}
\usepackage{amsmath}
\def\dlambda{$\lambda\lambda$}
\def\kms{km~s$^{-1}$} 

\slugcomment{Submitted to ApJ October 31, 2012; accepted June 7, 2013}

\shorttitle{3D Kinematics of Cas A}
\shortauthors{Milisavljevic \& Fesen}

\begin{document}

\title{A Detailed Kinematic Map of Cassiopeia A's Optical Main Shell\\
 and Outer High-Velocity Ejecta}

\author{Dan~Milisavljevic$^{1}$ and Robert A. Fesen$^{2}$}

\affil{$^{1}$Harvard-Smithsonian Center for Astrophysics, 60 Garden
  Street,
  Cambridge, MA, 02138\\
email: dmilisav@cfa.harvard.edu\\
  $^2$6127 Wilder Lab, Department of Physics \& Astronomy, Dartmouth
  College, Hanover, NH 03755}

\begin{abstract}

  We present three-dimensional kinematic reconstructions of optically
  emitting material in the young Galactic supernova remnant
  Cassiopeia~A (Cas~A).  These Doppler maps have the highest spectral
  and spatial resolutions of any previous survey of Cas~A and
  represent the most complete catalog of its optically emitting
  material to date.  We confirm that the bulk of Cas~A's optically
  bright ejecta populate a torus-like geometry tilted approximately
  30$\degr$ with respect to the plane of the sky with a $-4000$ to
  $+6000$ \kms\ radial velocity asymmetry. Near-tangent viewing angle
  effects and an inhomogeneous surrounding CSM/ISM environment suggest
  that this geometry and velocity asymmetry may not be faithfully
  representative of the remnant's true 3D structure or the kinematic
  properties of the original explosion. The majority of the optical
  ejecta are arranged in several well-defined and nearly circular
  ring-like structures with diameters between approximately
  $30\arcsec$ (0.5 pc) and $2\arcmin$ (2 pc). These ejecta rings
  appear to be a common phenomenon of young core-collapse remnants and
  may be associated with post-explosion input of energy from plumes of
  radioactive $^{56}$Ni-rich ejecta that rise, expand, and compress
  non-radioactive material. Our optical survey also encompassed
  Cas~A's faint outlying ejecta knots and exceptionally high-velocity
  NE and SW streams of S-rich debris often referred to as
  `jets'. These outer knots, which exhibit a chemical make-up
  suggestive of an origin deep within the progenitor star, appear to
  be arranged in opposing and wide-angle outflows with opening
  half-angles of $\approx 40\degr$.

\end{abstract}


\section{Introduction}
\label{sec:Intro}

A variety of observations and hydrodynamic modeling make a compelling
case that high-mass, core-collapse supernovae (SNe) are intrinsically
aspherical events with highly clumped ejecta caused by dynamical
instabilities \citep{Wang08,Maeda08,Nordhaus10,Janka12,Tanaka12}.  The
origin of the expansion asphericities is currently uncertain.
Possible causes include asymmetrical neutrino heating and
accretion-shock instabilities
\citep{Kifonidis00,Blondin03,Kifonidis03,Burrows06,Scheck06,MK09,Hanke12},
and the influences of rotation and magnetic fields
\citep{Wheeler02,Akiyama03,Fryer04,Shibata06,Masada12}.

Observations of the kinematic and chemical properties of SN ejecta can
help investigate which of the aforementioned explosion mechanisms may
dominate. For example, late-time optical spectra obtained $t \ga 6$
months beyond outburst in stripped-envelope events often exhibit
multi-peaked emission line profiles consistent with explosion models
simulating aspherical axisymmetric and potentially jet-related
explosions viewed along different angles from the equatorial plane
\citep{Mazzali05,Modjaz08,Maeda08,Taubenberger09,Milisavljevic10}. Additional
clues for constraining explosion mechanisms come from
spectropolarimetry. These studies show that ejecta can be asymmetric
in the inner layers, supporting the view that the explosion process is
strongly aspherical \citep{Wang08,Tanaka12}.

An alternative way to investigate core-collapse explosion asymmetries
is through observations of young Galactic supernova remnants
(SNRs). Studies of the handful of young O-rich Galactic remnants
believed to be the results of massive stars allow one to probe
kinematic asymmetries in the expanding ejecta at spatial scales not
possible from extragalactic SN observations.  They can also offer
clues about the explosive mixing of chemically distinct zones in the
progenitor star and the nature of the central compact remnant.

The young Galactic remnant Cassiopeia A (Cas~A) provides perhaps the
clearest look at the explosion dynamics of a high mass SN.  With an
explosion date most likely around $1681 \pm 19$, Cas~A is the youngest
Galactic core-collapse SNR known
\citep{Thorstensen01,Fesen06bowtie}. At an estimated distance of 3.4
kpc \citep{Reed95}, it is also among the closest.

Cas~A is the only historical core-collapse SNR with a secure SN
subtype classification. The detection of light echoes of the
supernova outburst \citep{Rest08,Rest11,Besel12} has enabled
follow-up optical spectral observations which indicate that the
original supernova associated with Cas~A exhibited an optical spectrum
at maximum light similar to those seen for the Type IIb events SN~1993J
and SN~2003bg \citep{Krause08,Rest11}.

Cas~A is inferred to have undergone extensive mass loss from its
original $20-25$ M$_{\odot}$ progenitor to only $3-4$\,M$_{\odot}$
upon explosion, leaving a relatively dense and slow moving remnant
stellar wind \citep{Chevalier03Clumpy}. Such conditions do not
generally arise with a radiatively driven wind from a $20-25$
M$_{\odot}$ progenitor, but require the existence of a binary
companion to aid the mass loss
\citep{Woosley93,Young06,Claeys11}.

Several Doppler reconstructions of Cas~A's main shell ejecta have been
conducted in the past using optical, infrared, and X-ray data. Results
of these studies have revealed significant ejecta asymmetries
potentially related to the explosion dynamics. For example, optical
studies have shown that much of the remnant's ejecta are arranged in
large rings on a spherical shell exhibiting an overall velocity
asymmetry of $-4000$ to $+6000$ \kms\
\citep{Minkowski59,Lawrence95,Reed95}. Spectra of surrounding light
echos also indicate strong asymmetry in the Cas\,A supernova's
photosphere \citep{Rest11}.

\begin{figure*}

\centering
\includegraphics[width=0.85\linewidth]{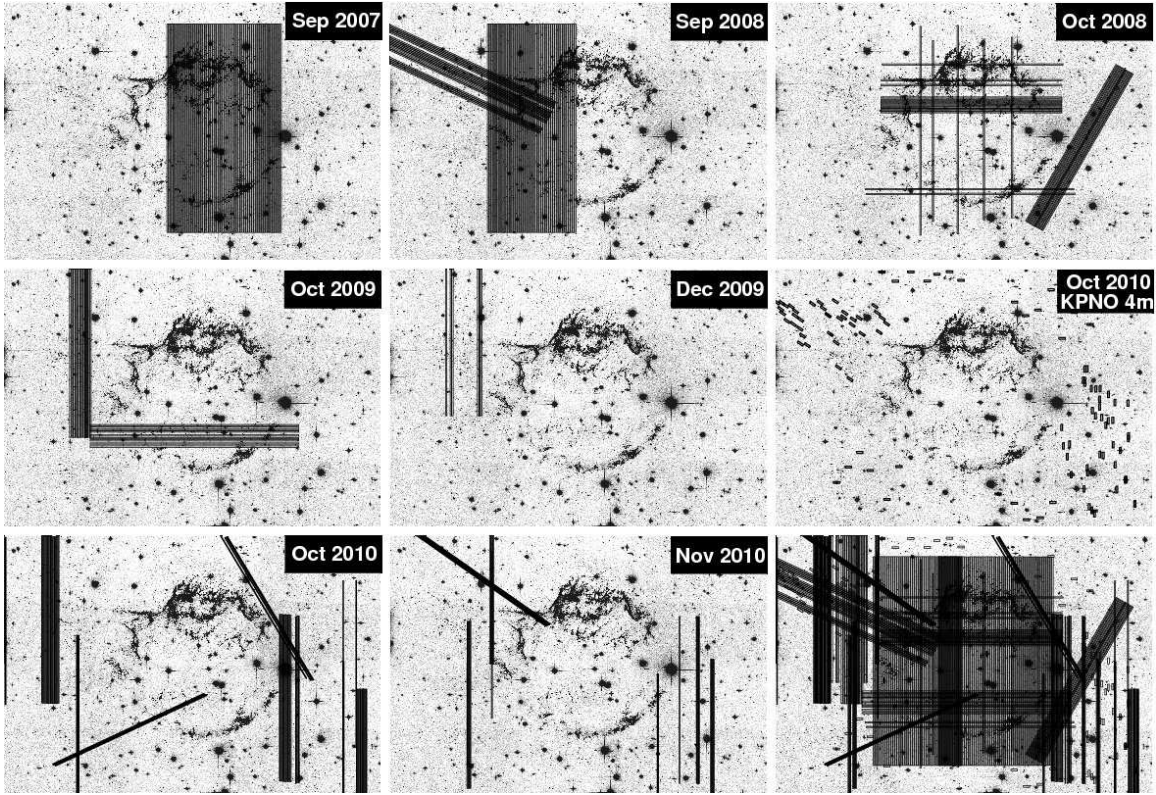}

\caption{Finding charts of all long-slit positions. Background image
  is a mosaic created from 2004 {\sl HST}/ACS observations sensitive
  to oxygen and sulfur emissions (Proposal 10286; PI: R.\ Fesen; see
  \citealt{Fesen06a}). Bottom right is montage of all locations. Refer
  to Table~\ref{tab:ObsSummary} for telescope and instrument details.}

\label{fig:CasAslitmap_all}
\end{figure*}

\begin{deluxetable*}{llll}
\footnotesize
\centering
\tablecaption{Summary of Observations}
\tablecolumns{4}
\tablewidth{0pt}
\tablehead{\colhead{Date}                &
           \colhead{Telescope}           &
           \colhead{Instrument}          &
           \colhead{Region (No. of Positions)} }
\startdata
2007 Sep 01-08    & MDM 2.4 m & Modspec & Main Shell (58)               \\
2008 Sep 18-25    & MDM 2.4 m & Modspec & Main Shell (45); NE Jet (18)  \\
2008 Oct 27-30    & MDM 2.4 m & MKIII   & Main Shell (19); SW Jet (10)  \\
2009 Oct 19-22    & MDM 2.4 m & CCDS    & Main Shell (9); NE Jet (16)   \\
2009 Dec 11-14    & MDM 2.4 m & CCDS    & NE Jet (6)                    \\
2010 Oct 1-5      & KPNO 4 m  & MARS    & NE/SW Jets + Outer Knots (81) \\
2010 Oct 8-13     & MDM 2.4 m & MKIII   & NE/SW Jets + Outer Knots (38) \\
2010 Oct 28-Nov 1 & MDM 2.4 m & MKIII   & NE/SW Jets + Outer Knots (26) \\ 
\enddata
\label{tab:ObsSummary}
\end{deluxetable*}

The recent study of the Cas~A remnant by \citet{DeLaney10} presented
an extensive multi-wavelength Doppler reconstruction using new
velocity measurements from {\sl Spitzer} infrared and {\sl Chandra}
X-ray observations combined with previous optical data on the
remnant's highest velocity, outer ejecta from \citet{Fesen01Turb}.
Besides confirming the presence of several large ejecta rings in the
infrared, they interpreted the structure of Cas~A's bright main shell
of ejecta knots to consist of a spherical component, a tilted thick
disk, multiple ejecta jets/pistons, and optical fast-moving knots all
populating the thick disk plane. They concluded that the bulk of the
symmetries and asymmetries seen in Cas A are properties intrinsic to
the supernova explosion.

One often cited piece of observational evidence that the Cas~A
progenitor underwent a highly aspherical explosion has been the
presence of a `flare' or `jet' of unusually high-velocity SN debris
extending some $300\arcsec$ out along the northeast (NE) limb from the
remnant center (about twice that of main shell ejecta) at a position
angle of $\sim 70\degr$ and visible even in the earliest photographic
images of the remnant \citep{Minkowski68,vandenBergh70}.  Ejecta knots
in this NE region exhibit proper motions indicating expansion
velocities extending up to $14,000$ (D/3.4 kpc) \kms, some $8000$ \kms
\ faster than the fastest main shell ejecta knots
\citep{Fesen01Turb,Fesen06bowtie}.

A fainter and considerably sparser southwest (SW) so-called
`counterjet' of equally high-velocity ejecta was discovered optically
\citep{Fesen01Turb} and subsequently confirmed in X-rays and the
infrared \citep{Hwang04,Hines04}.  {\sl Chandra} X-ray images show the
SW jet to be much fainter and not as extended radially as the NE one,
and in the optical only about $200$ knots have been identified
compared to over $1000$ in the NE jet. One reason for this substantial
difference, at least in the optical, may lie in greater line of sight
extinction (A$_{V} \approx 5-8$ mag) suspected along the remnant's
western boundary \citep{Keohane96,Willingale02}.

\begin{figure*}
\centering
\includegraphics[width=0.80\linewidth]{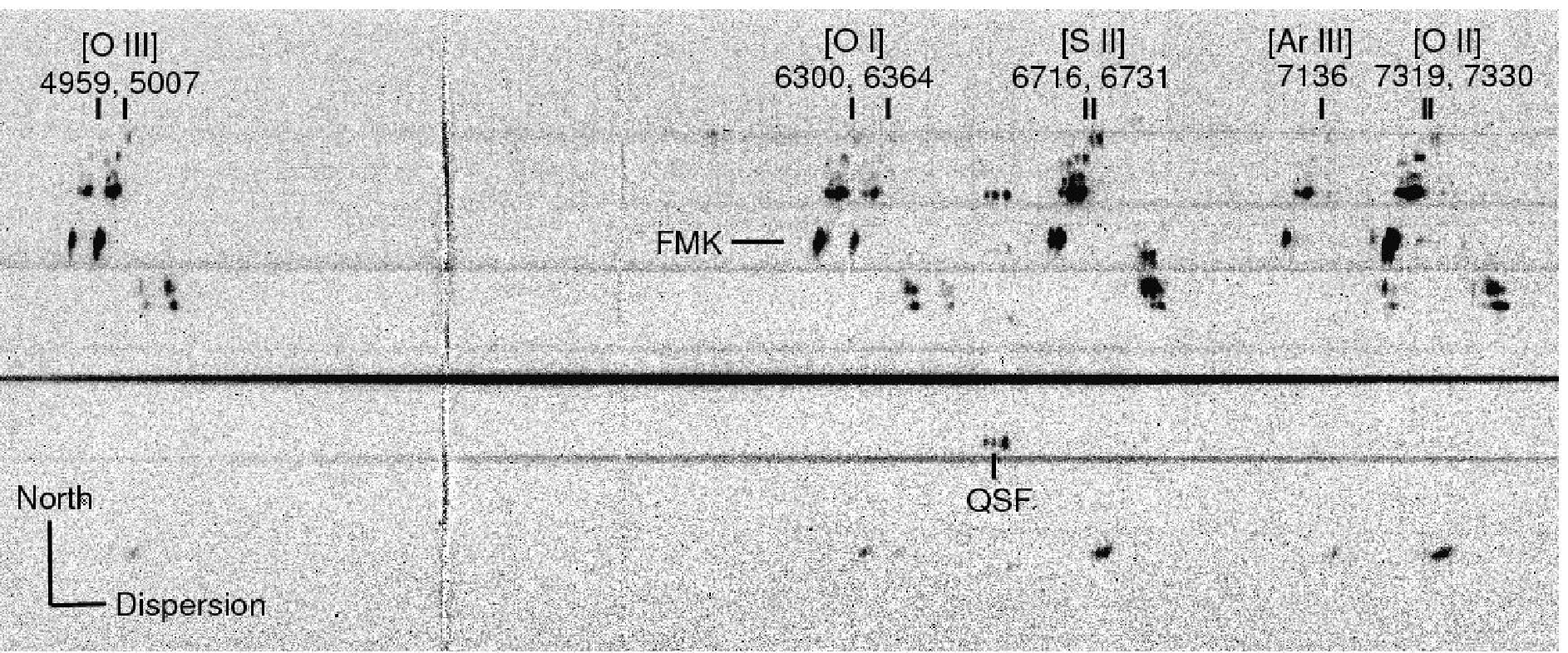}

\caption{Example of a fully reduced and cleaned 2D spectrum of a
  slit position along the main shell from which 1D extractions were
  made. Main shell fast-moving knots (FMKs) and a CSM-related
  quasi-stationary material (QSF) are labeled. Wavelength positions
  for the brightest emission lines are marked.}

\label{fig:2Dspec}
\end{figure*}

Some theoretical models have suggested that bipolar jets may be
associated with the explosion dynamics. For example, numerical
simulations of MHD jet and neutrino-driven expansion models produce
aspherical jets with ejection velocities and axial expansion ratios
$\sim 2$ not unlike those seen for Cas~A
\citep{Khokhlov99,Kotake05}. While an X-ray analysis of Cas~A's NE and
SW jets concluded that they are unlikely to have played an important
role in the explosion mechanism \citep{Laming06}, the kinematic
properties of the NE and SW jets have not been well determined and the
limited data currently available for these regions cannot constrain
whether or not the two streams even comprise a genuine bipolar,
jet-counterjet expansion structure. Such information is crucial to
assess their potential relationship to core-collapse explosion
dynamics.

The need for an in-depth study of the kinematic properties of the NE
and SW ejecta jets and their relation to the remnant's main shell of
reverse shock heated ejecta motivated us to undertake a deep
reconnaissance of the entire Cas~A remnant with high spatial and
spectral resolution. In Sections~\ref{sec:Observations} and
\ref{sec:Data}, we describe the observations and our methods for
reducing the data. This is followed with a brief description of how
these data were used to develop a three dimensional (3D) Doppler
reconstruction of the optically emitting ejecta in
Section~\ref{sec:Doppler}. In Sections~\ref{sec:Results} and
\ref{sec:Discussion}, the kinematic properties of the ejecta are
presented and discussed. We summarize our findings and discuss
potential future work in Section~\ref{sec:Conclusions}.

\section{Observations}
\label{sec:Observations}

A series of observing runs starting September 2007 and continuing to
November 2010 were conducted to obtain low-dispersion long-slit
optical spectra across the entire remnant. Most observations were
carried out at MDM Observatory, on Kitt Peak, AZ using the 2.4m
Hiltner telescope. These observations were later supplemented by both
multi-slit and long-slit spectra obtained with the Kitt Peak Mayall 4m
telescope. A summary of all observations is given in
Table~\ref{tab:ObsSummary}. Additional relevant details of these
observations are discussed below. A finding chart of all slit
positions is presented in Figure~\ref{fig:CasAslitmap_all}.

For Cas~A's main shell ejecta, the MDM Modular Spectrograph (Modspec)
with a SITe $2048 \times 2048$ CCD detector (`Echelle') on the MDM 2.4
m telescope was used. A $2\arcsec \times 5\arcmin$ long-slit oriented
north--south was spaced every 3$\arcsec$.  Exposures were generally $2
\times 500$~s. The effective wavelength coverage was approximately
$4500-7700$~\AA\ with a spectral resolution of 6 \AA.  Comparison
Ne--Hg--Xe lamps were taken every five positions to track instrument
flexure. To simplify the observations, long-slits were generally not
orientated along the parallactic angle. However, all observations were
obtained at airmass $< 1.4$ to limit the amount of light lost to
atmospheric differential refraction.

\begin{figure*}
\centering
\includegraphics[width=0.425\linewidth]{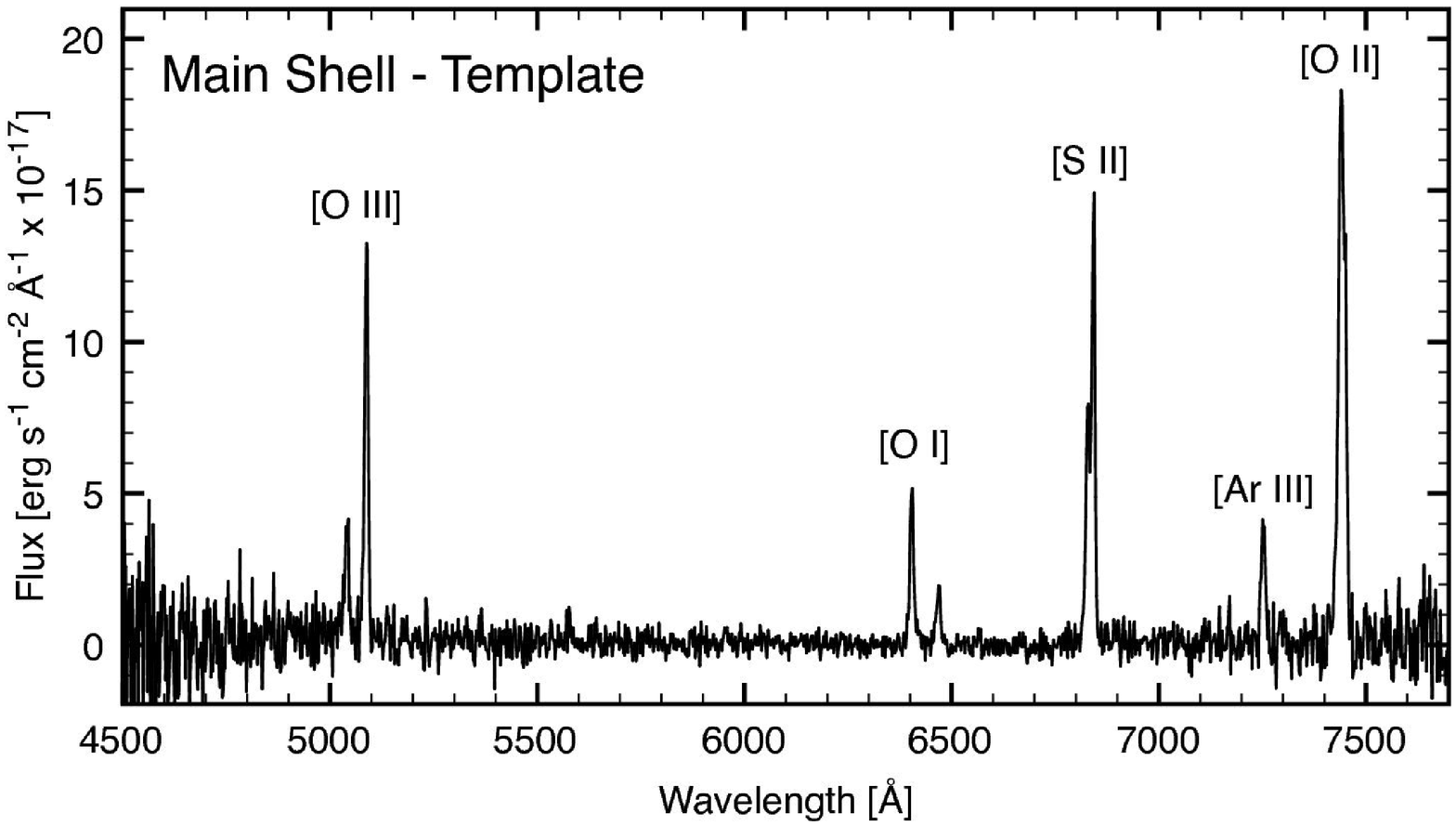}
\includegraphics[width=0.425\linewidth]{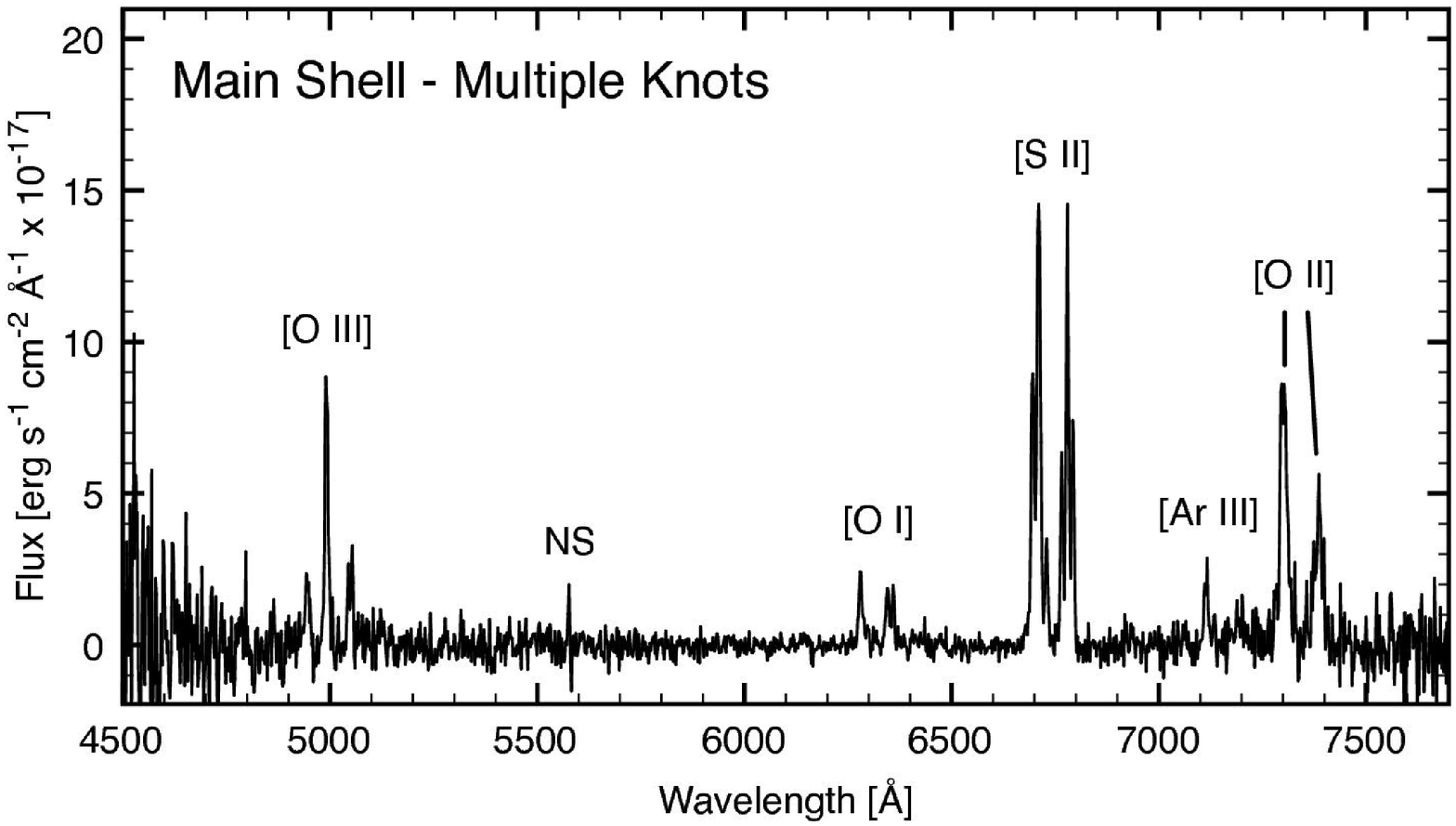}\\
\includegraphics[width=0.425\linewidth]{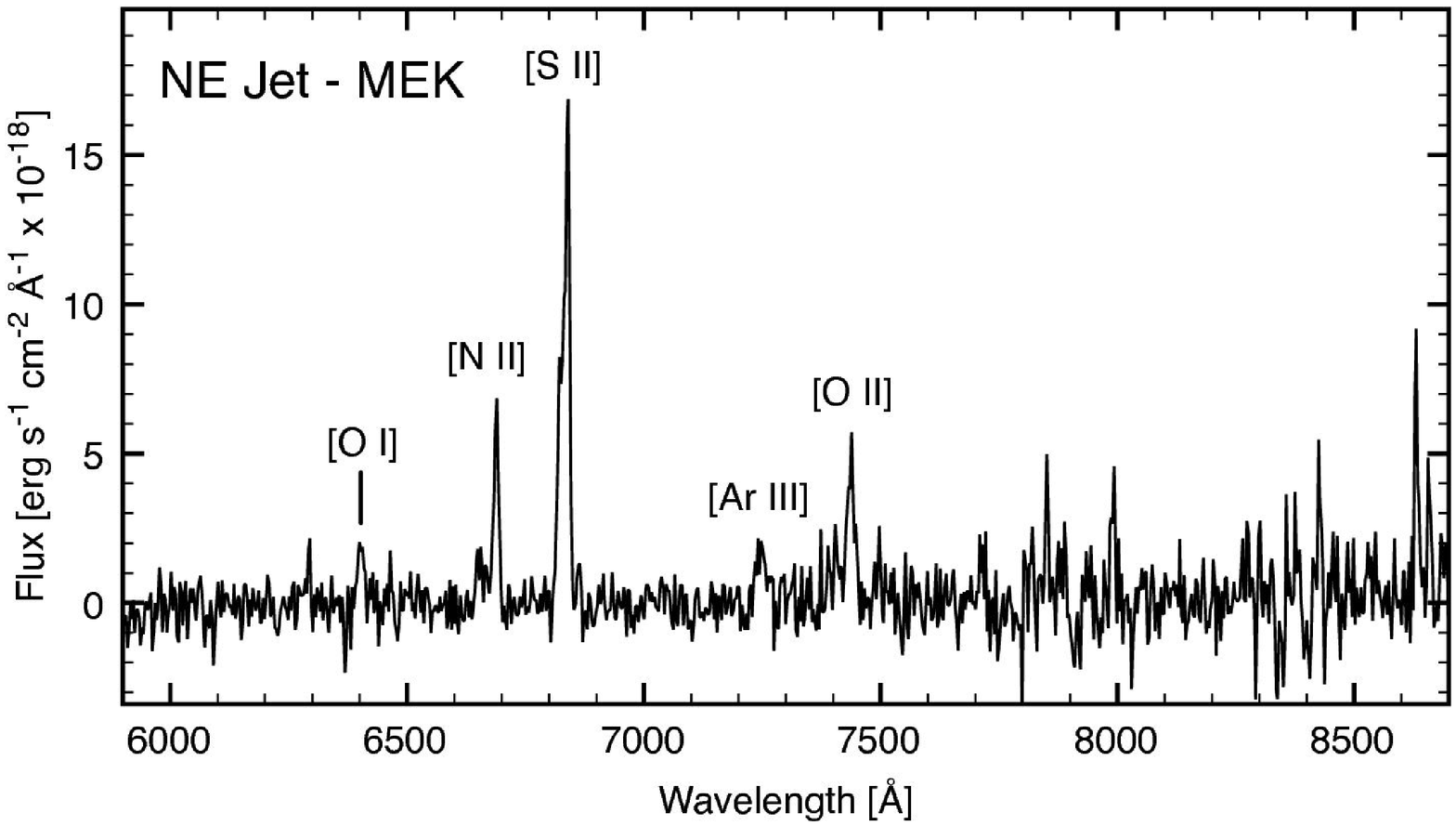}
\includegraphics[width=0.425\linewidth]{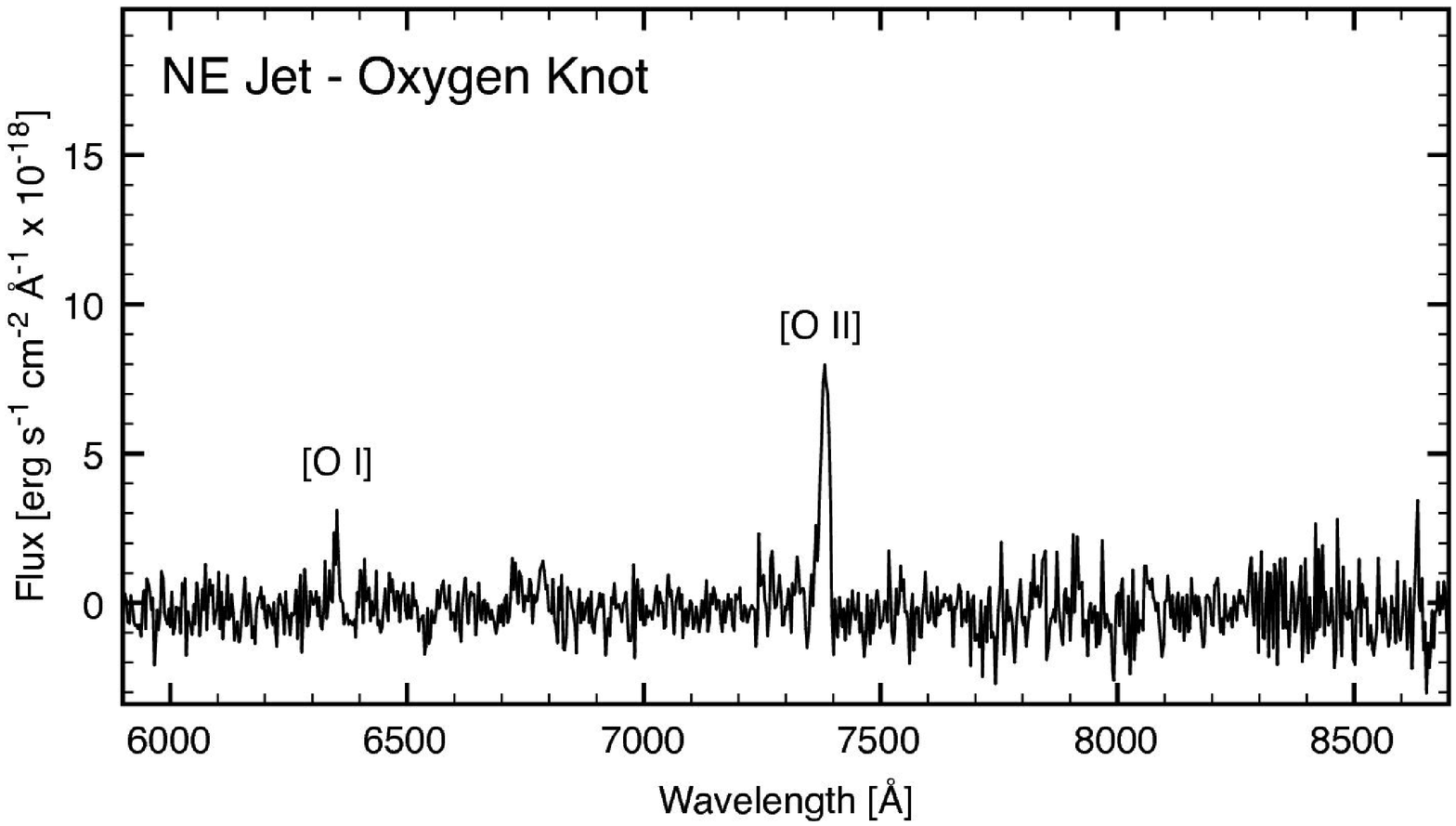}

\caption{Examples of 1D spectra of ejecta knots of Cas~A used in the
  Doppler reconstruction. The top left spectrum from the main shell is
  an example template used to identify knots and determine radial
  velocities. Top right shows a spectrum with multiple overlapping
  knots. Bottom left is a Mixed Ejecta Knot (MEK) from the NE Jet
  region. Bottom right is an oxygen-rich knot also from the NE Jet
  region.}

\label{fig:1Dspec}
\end{figure*}

For the remnant's NE and SW jets, along with some outer ejecta, a variety of
instruments on the MDM 2.4 m were used. Some data were obtained using the
Mark~III Spectrograph (MKIII) employing a $1\farcs 7 \times 4.5\arcmin$ slit
and using a SITe $1024 \times 1024$ CCD detector (`Templeton'). A 300 lines
mm$^{-1}$ 5400 \AA\ blaze grating yielded spectra spanning $4500-7400$ \AA\
with 10~\AA\ resolution. Other data were obtained using the Boller and Chivens
CCD Spectrograph (CCDS) with a $2\arcsec \times 4.5\arcmin$ slit and either the
(i) blue-sensitive 150 line mm$^{-1}$ 4700 \AA\ blaze or (ii) red sensitive 158
line mm$^{-1}$ 7530 \AA\ blaze grating. Both of these setups had approximately
12 \AA\ resolution. Comparison lamps and flats were taken to track telescope
flexure and aid in the subtraction of fringing. These long-slit spectra were
taken using longer exposures of $1000-2000$ s. Slits were oriented to maximize
the number of ejecta knots detected at each slit position.

Standard stars from \citet{Stone77} and \citet{Massey90} were observed
each night. Comparison between evenings of each run generally showed
good agreement, and for the main shell observing runs all standards
were combined into one sensitivity function from which data were
calibrated for relative flux. Some nights, particularly those for the
outer ejecta knots, showed evidence of variable conditions and
were flux calibrated independently on a night-by-night basis.

Multi-slit spectra of both NE and SW jet knots and other outlying
ejecta were also obtained October 2010 with the Mayall 4m. The
Multi-Aperture Red Spectrometer (MARS) was used in combination with
the red-sensitive LBNL CCD detector and VPH 8050-450 grism. Spectra
covered the wavelength region of $5500-10,8000$ \AA\ with resolution of
$\approx 8$~\AA. A total of 12 slit masks were made to cover 81
knots. We had 24 successful detections. Comparison He-Ne-Ar lamp
images were taken for wavelength calibration, and standard star
observations consistent with those made for the MDM runs were used for
flux calibration.

\section{Data Reduction}
\label{sec:Data}

Data were reduced homogeneously through a series of scripts written to
automate \texttt{IRAF/PyRAF}\footnote{IRAF is distributed by the
  National Optical Astronomy Observatories, which are operated by the
  Association of Universities for Research in Astronomy, Inc., under
  cooperative agreement with the National Science Foundation. PyRAF is
  a product of the Space Telescope Science Institute, which is
  operated by AURA for NASA.}  procedures.  The 2D images of each
position were first trimmed, bias-subtracted, flattened, and co-added
to remove cosmic rays. Images were then wavelength calibrated in the
dispersion axis using the comparison lamp images and straightened in
the spatial axis using tracings of stellar continua.

\begin{figure}[h!tf]
\centering
\includegraphics[width=0.9\linewidth]{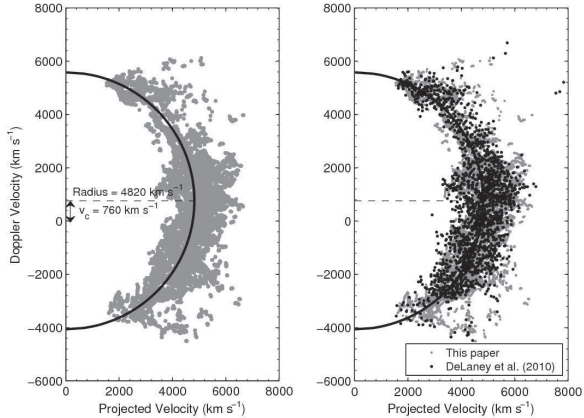}

\caption{Measured main shell knot velocities. Projected radii from the
  COE have been converted to velocities using scale factor $S$ = 0.022
  \kms arcsec$^{-1}$. {\it Left}: Data from this paper. Best fit
  semi-circle is shown. {\it Right}: Data from \citet{DeLaney10} for
  infrared \ion{Ar}{2} emissions shown over top data from this paper.}

\label{fig:Dopplerfit}
\end{figure}

Excellent background subtraction of the images was achieved through
the IRAF task \texttt{background} using a fifth-order chebyshev
function fit along the spatial direction that was sampled from the
median of 50 pixel bins. This provided sky emission clean 2D images
from which 1D spectra could be extracted. An example of a reduced and
cleaned 2D image is shown in Figure~\ref{fig:2Dspec}.

Each line of the 2D spectra was extracted and added in weighted
triplet groupings.  No effort was made to catalog the H$\alpha$,
[\ion{N}{2}] \dlambda 6548, 6583, and \ion{He}{1} emissions associated
with circumstellar material (CSM) often referred to as
Quasi-Stationary Flocculi (QSF; \citealt{vandenBergh70}) frequently
encountered (see Fig.~\ref{fig:2Dspec}). Investigation of this pre-SN
mass loss material is left for future work.

From these 1D extractions, knot velocities were measured through
cross-correlation of templates using the task \texttt{xcsao}. The
template was based on actual data from a representative knot
exhibiting the common lines of [\ion{O}{3}] \dlambda 4959, 5007,
[\ion{O}{1}] \dlambda 6300, 6364, [\ion{S}{2}] \dlambda 6716, 6731,
[\ion{Ar}{3}] $\lambda$7136, and [\ion{O}{2}] \dlambda 7319, 7330. We
show example spectra of four types of ejecta knots encountered in
Figure~\ref{fig:1Dspec}.  \texttt{xcsao} was run on all extracted 1D
spectra for Doppler velocities running between $-6500$ to $+6500$
\kms\ to find individual knots. This procedure was found to work well
even in cases with multiple knots (Figure~\ref{fig:1Dspec}, top
right).

Results of the preliminary selection were then manually inspected for
false positives. The remnant's faint jet and other outlying ejecta
knots required additional care owing to the range of emission lines
expressed, such as mixed emission knots (Figure~\ref{fig:1Dspec},
bottom left), and oxygen-rich knots (Figure~\ref{fig:1Dspec},
bottom right). See \citet{Fesen96} for additional details about the
spectroscopic properties of these types of knots. In some cases
velocities were determined manually using de-convolving techniques
available in the task \texttt{splot}.

Radial velocities of individual knots were then assigned coordinates
in right ascension (RA) and declination (DEC). A variety of
cross-checks ensured consistent and accurate coordinates. Stars with
well-measured coordinates encountered during the progression of
long-slit positions provided fudicial reference points. In some cases,
bright stars were intentionally observed to provide stellar continua
by which accurate positional offsets could be determined. Long-slit
positions oriented east-west were checked against overlapping slits
oriented north-south to further check positions and flux levels. A
final check to all positions was aided through comparison with
high-resolution {\sl Hubble Space Telescope} ({\sl HST}) images
(Proposal 10286; PI: R.\ Fesen). Measured velocities are believed
accurate to $\pm 40$ \kms\ at the 68\% confidence level. Positional
uncertainties are estimated to be no more than $2\arcsec$; that is, of
order the slit width.

\section{3D Doppler Reconstruction}
\label{sec:Doppler}

Cas~A's relatively young age and nearby distance make it a prime
candidate for Doppler velocity reconstruction. Previous studies show
that its ejecta knots lie roughly on a spherical shell traveling
radially outward from a unique center of expansion (COE;
\citealt{Reed95,Lawrence95,DeLaney10}). The COE is known to within one
arcsecond precision and lies at the coordinates $\alpha(2000.0) =
23^{\rm h}23^{\rm m}27\fs77$ $\delta(2000.0) = +58\degr48\arcmin
49\farcs4$ \citep{Thorstensen01}. A lack of detectable deceleration in
the proper motion of these knots over the last 50~yr and velocity
changes of less than 5\% over 300 yr
\citep{vandenBergh70,Kamper76,Thorstensen01} allows one to assume
ballistic trajectories from the COE described by the relation
\begin{equation} v = r
\times S, \label{eq:scale} 
\end{equation}
where $v$ is the radial velocity, $r$ is the angular distance 
from the explosion center, and $S$ is the scaling factor.

The value of $S$ can be found by fitting the measured Doppler
velocities to a spherical expansion model. We follow procedures
reviewed in \citet{DeLaney10}, which itself closely followed the work
of \citet{Reed95}. The model is a semi-circle of the velocity
distribution, which can be parameterized in terms of the center of the
velocity distribution, $v_c$, the minimum velocity at which the
semi-circle crosses the velocity axis, $v_m$, and $S$, which relates
the velocity axis to the spatial axis as:

\begin{equation}
(r_{p}/S)^2 + (v_{D} - v_{c})^2 = (v_{c} - v_{m})^2,
\end{equation}

\noindent where $r_{p}$ is the observed projected radius and $v_{D}$
is the observed Doppler velocity. 

In Figure~\ref{fig:Dopplerfit}, the results of our least-squares fit
to the data are shown. The calculated Doppler velocities are $v_{c} =
760 \pm 100$ \kms, $v_{m} = -4060 \pm 200$ \kms, and the scaling
factor is $S = 0\farcs 022 \pm 0\farcs 003$ per \kms. The results are
consistent with those determined by \citet{Reed95} using optical data,
as well as with \citet{DeLaney10} using {\it Spitzer} IR data of
[\ion{Ar}{2}] emission. Knot positions in RA and DEC were scaled to
velocities using Equation \ref{eq:scale}. We used the COE and the
calculated Doppler velocity $v_{c}$ to define a three-dimensional
center of expansion in velocity space from which all reported vector
trajectories originate.

\begin{figure*}
\centering

\includegraphics[width=0.55\linewidth]{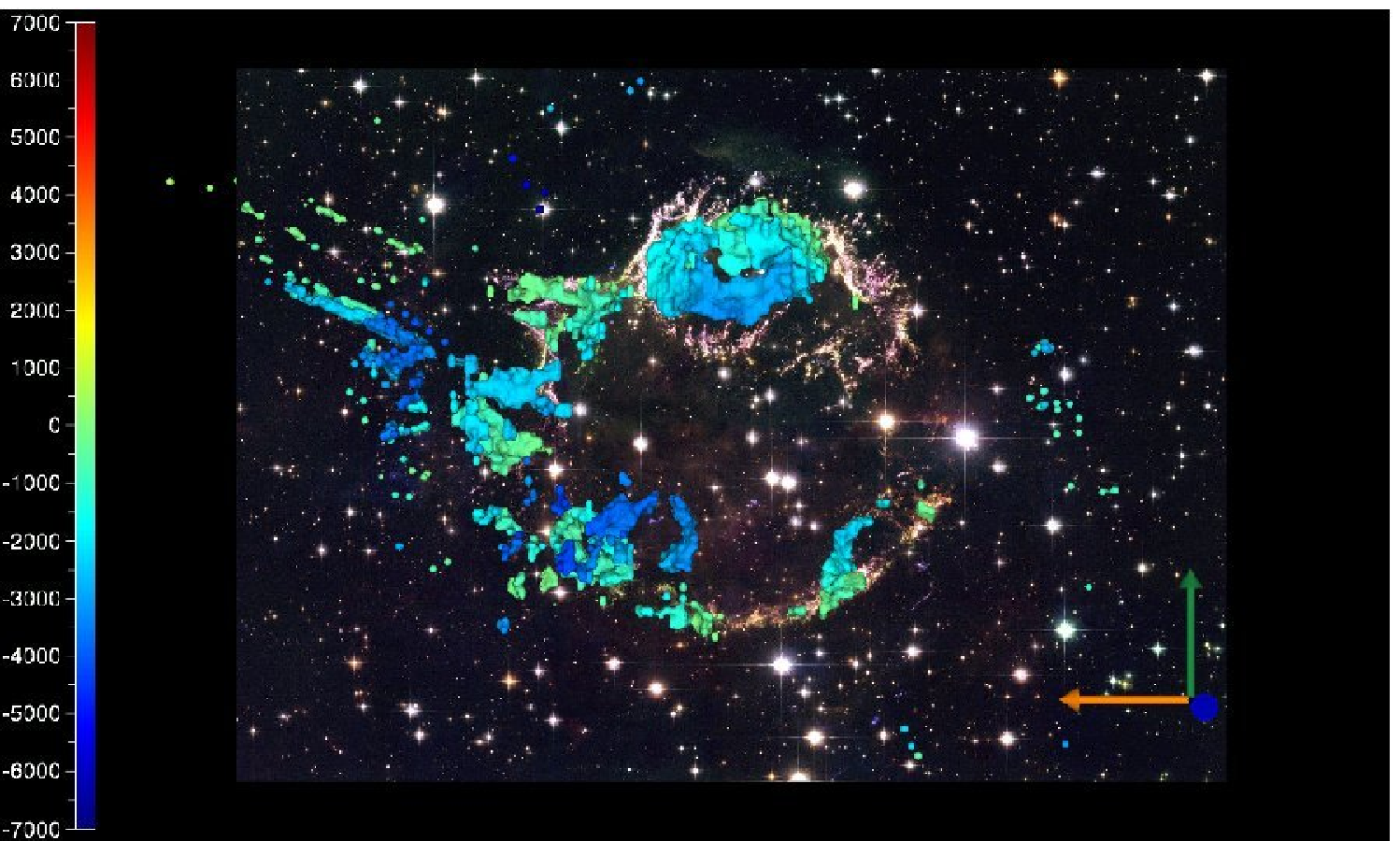}\\
\includegraphics[width=0.55\linewidth]{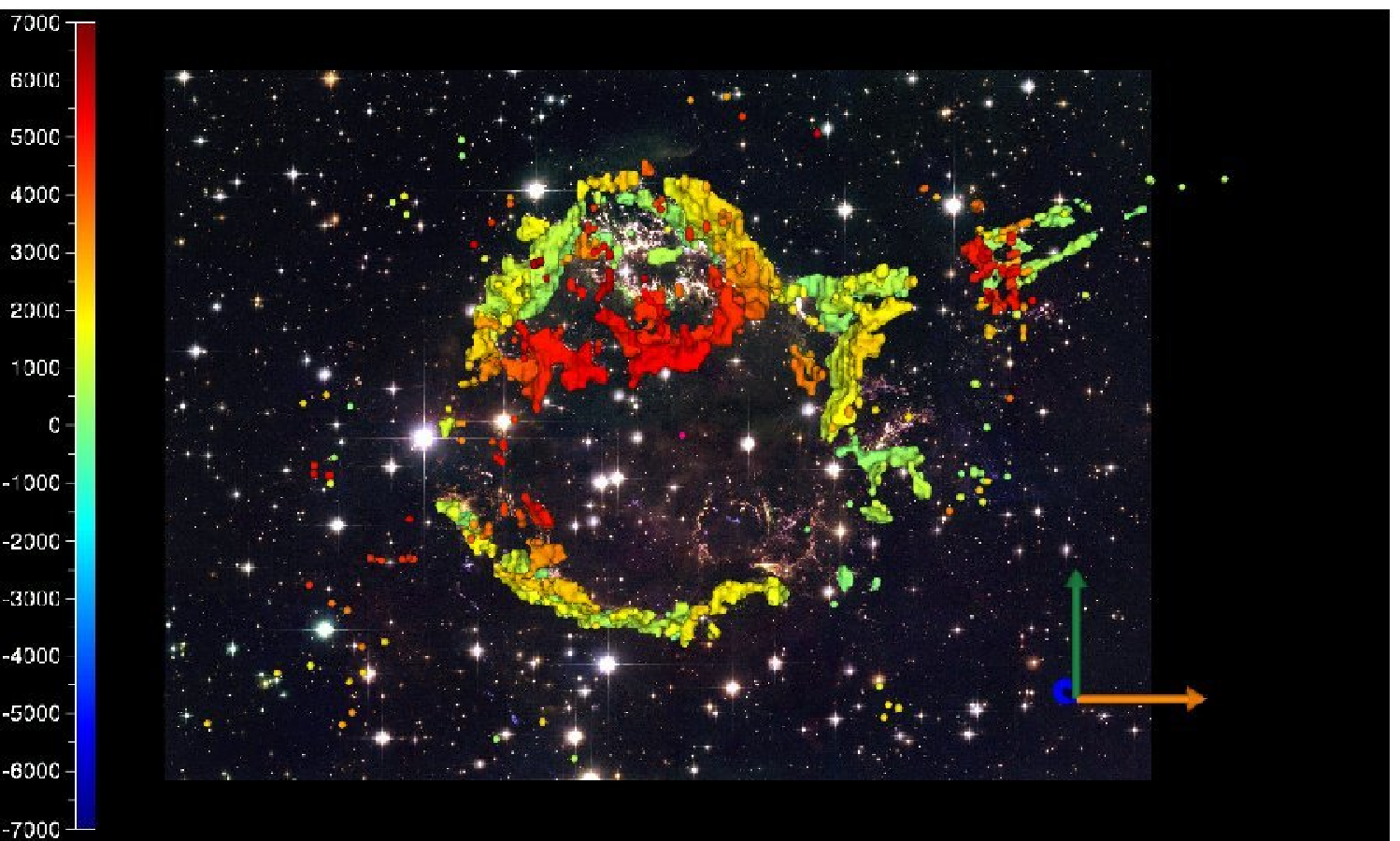}\\
\includegraphics[width=0.55\linewidth]{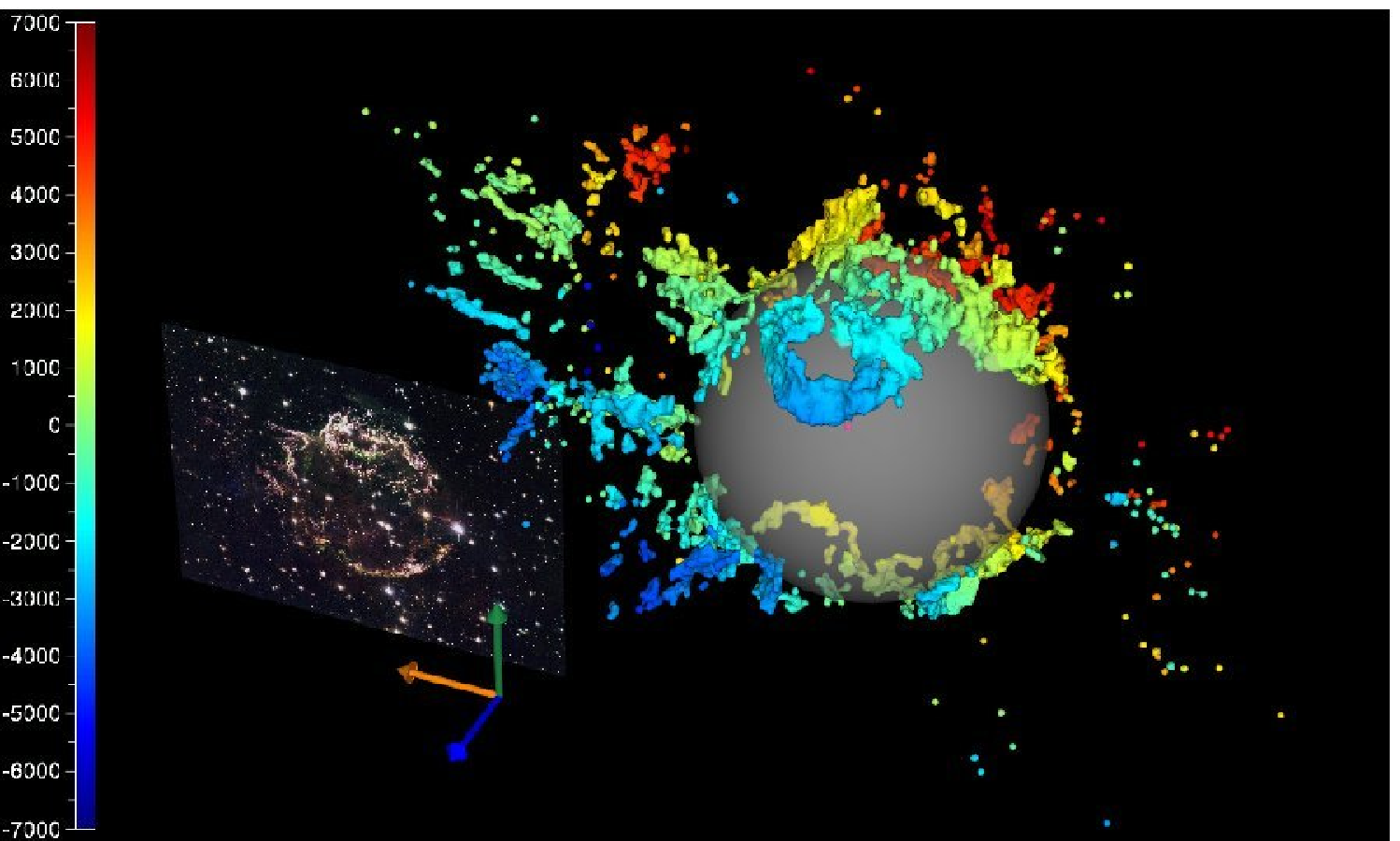}

\caption{3D Doppler reconstruction of Cas~A's optically emitting
  ejecta. The main shell and all sampled high velocity outer material
  is represented. Top panel shows the reconstruction with respect to
  the plane of the sky as observed from Earth with north up and east
  to the left. The velocity gradient is color-coded and only
  blueshifted material is visible from this perspective. Middle panel
  shows the same perspective rotated 180$\degr$ with respect to the
  north-south axis. In this representation, the vantage point is from
  behind Cas~A with only redshifted material shown. Bottom panel is an
  angled perspective showing the full range of velocities. The plane
  of the sky is shown offset for reference. A translucent sphere is a
  visual aid to help distinguish between front and back
  material. Refer to Movie 1 for an animation of these data.}

\label{fig:CasA3D}
\end{figure*}

\section{Results}
\label{sec:Results}

The results from our survey encompassing 13,769 individual data points
are presented in Figure~\ref{fig:CasA3D}.  Incorporated into the data
set are an additional 73 outer ejecta knots with measured radial
velocities reported in \citet{Fesen01Turb}. The resulting Doppler maps
have the highest spectral and spatial resolutions of any previous
survey of Cas~A, and represent the most complete catalog of its
optically emitting ejecta material to date.

\begin{figure*}[htp!]
\centering

\includegraphics[width=0.4\linewidth]{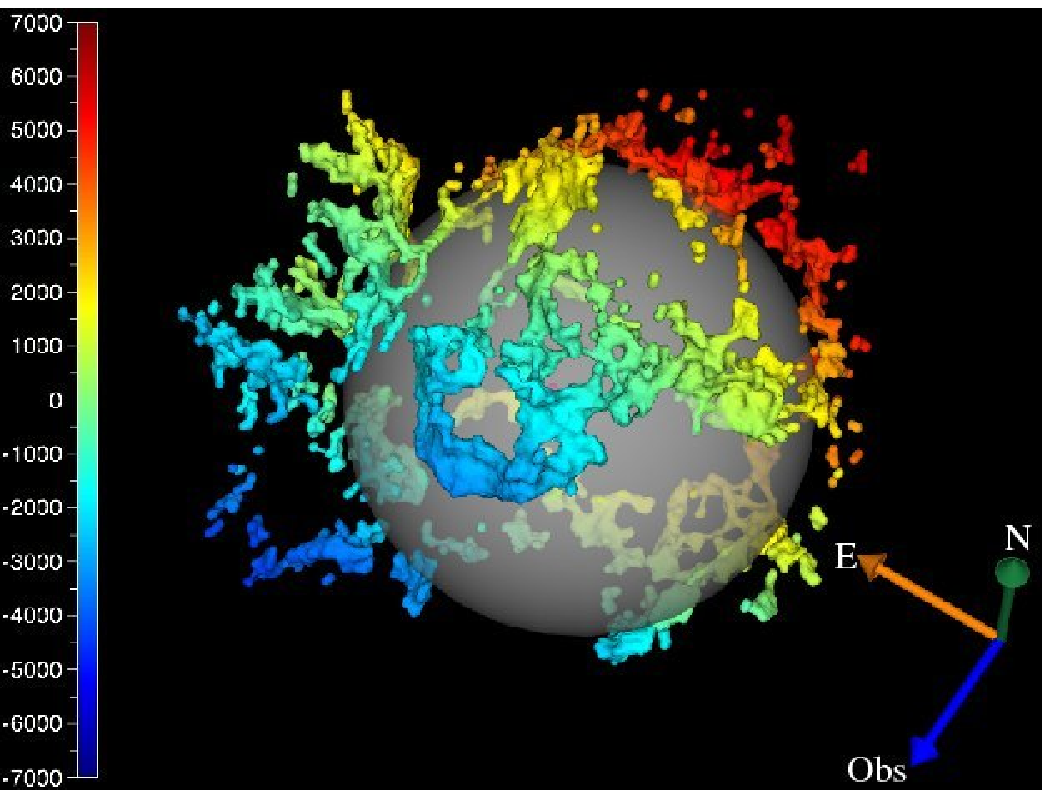}
\includegraphics[width=0.4\linewidth]{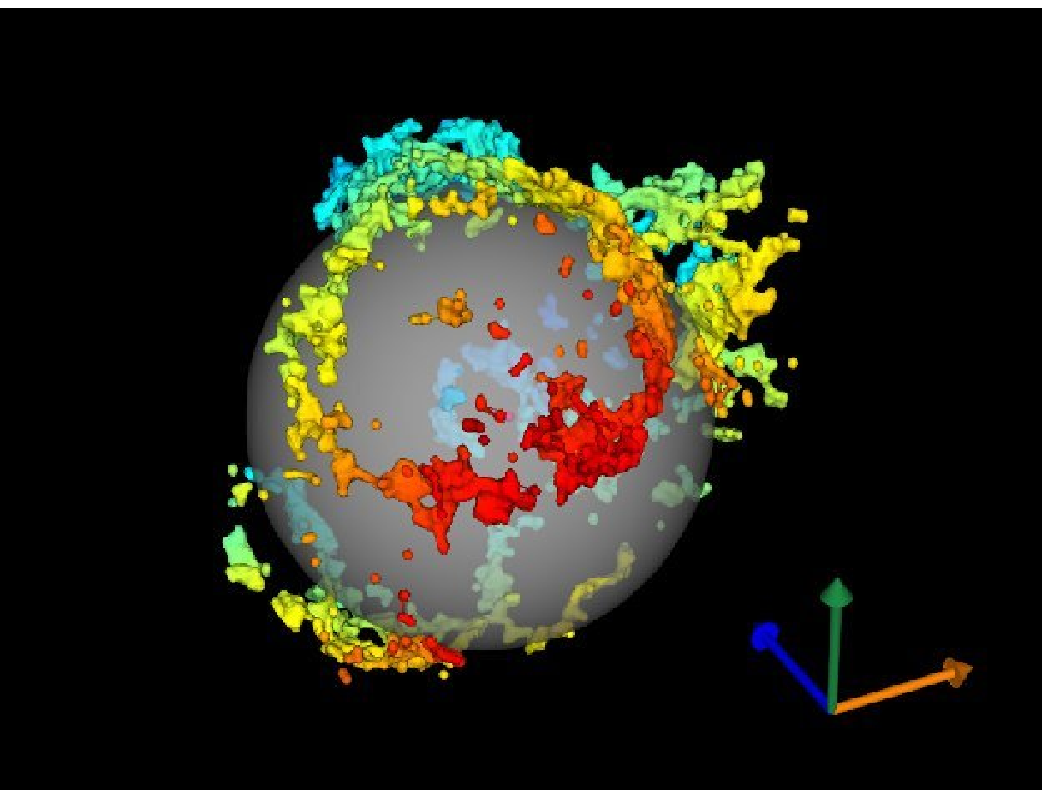}

\vspace{0.05cm}

\includegraphics[width=0.4\linewidth]{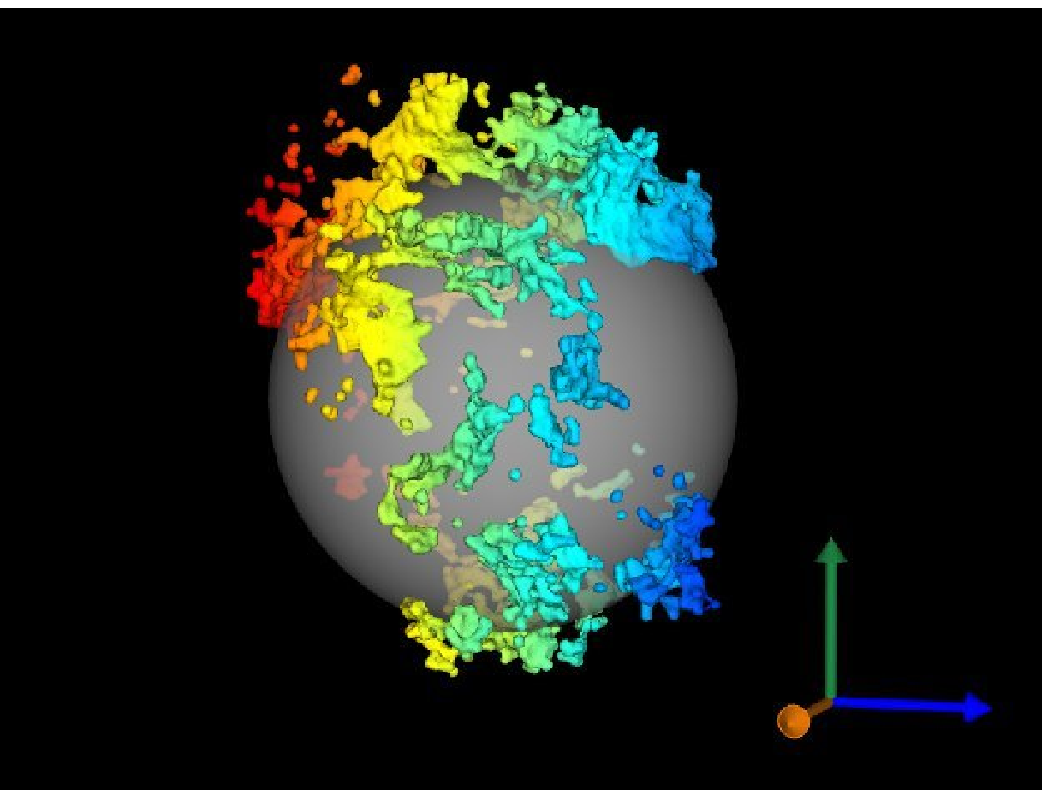}
\includegraphics[width=0.4\linewidth]{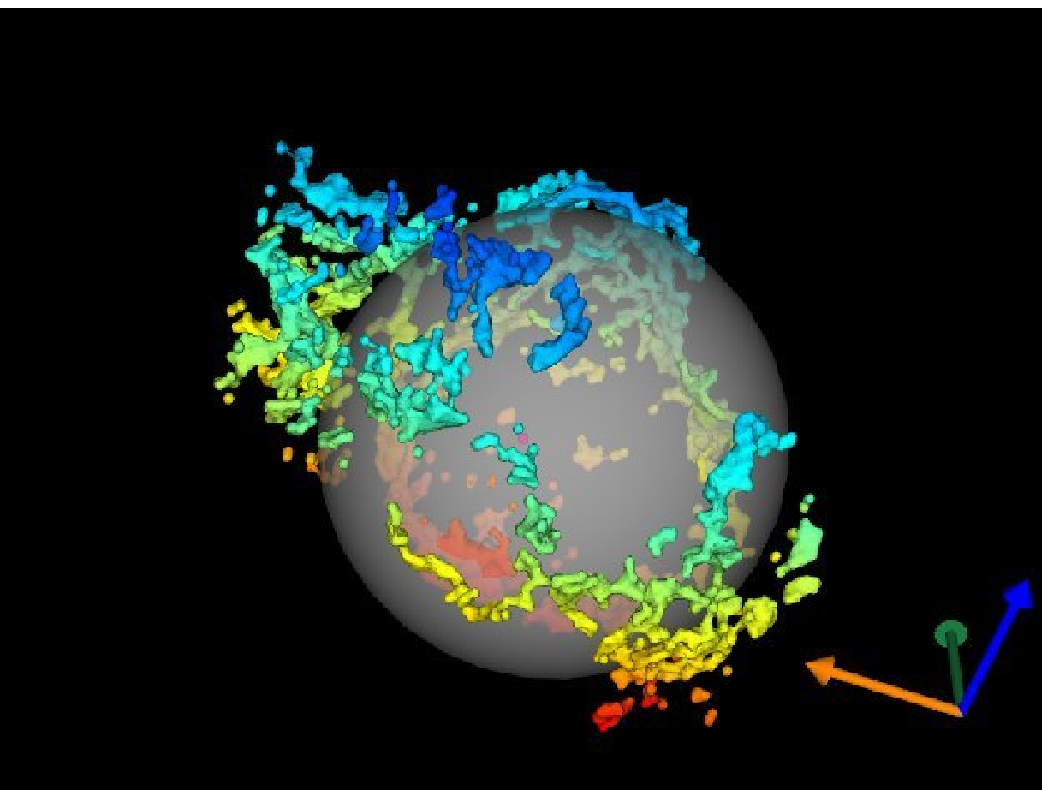}

\caption{Various perspectives of the Cas~A reconstruction illustrating
  the ubiquity of large ejecta rings.  Only ejecta knots in the main
  shell are shown. Top left panel shows some of the more prominent
  ejecta structures. Noticeable rings of ejecta include the
  blueshifted ring in the north, and the much larger neighboring
  redshifted ring. Also seen is the extent to which the rings are
  extended radially and hence more crown-like than simple flat
  rings. The top right panel shows a better perspective of the large
  northern redshifted ring. Bottom left panel shows a side profile
  viewing towards the base of the NE jet, while the bottom right panel
  shows Cas~A from below where a southwestern ring can be seen as well
  as a number of partial or broken ring structures. Refer to Movie 1
  for an animation of these data.}

\label{fig:Rings}
\end{figure*}

Measured knot radial velocities are represented in
Figure~\ref{fig:CasA3D} with a color-coded gradient. A surface
reconstruction has been performed using a ball-pivoting algorithm to
interpolate and smooth the original point cloud.  The Doppler map is
plotted against a background image of Cas~A created from the {\sl HST}
ACS/WFC images (GO 10286; PI Fesen) obtained in filters F625W (colored
blue), F775W (green), and F850LP (red) as a visual aid to relate the
3D kinematic structure to Cas~A's appearance on the plane of the sky.
Three viewing orientations are shown: a face-on view, a rear view, and
an angled view from above shown with the background {\sl HST} image
offset.  An animation has also been created showing the individual
data points and the surface reconstruction, rotated along the
north-south and east-west axes, with and without the outer
high-velocity ejecta knots (Movie 1).

Below we discuss some specific properties of Cas~A highlighting our
high-resolution 3D reconstructions of the remnant's kinematic
structure.

\begin{figure}
\centering
\includegraphics[width=\linewidth]{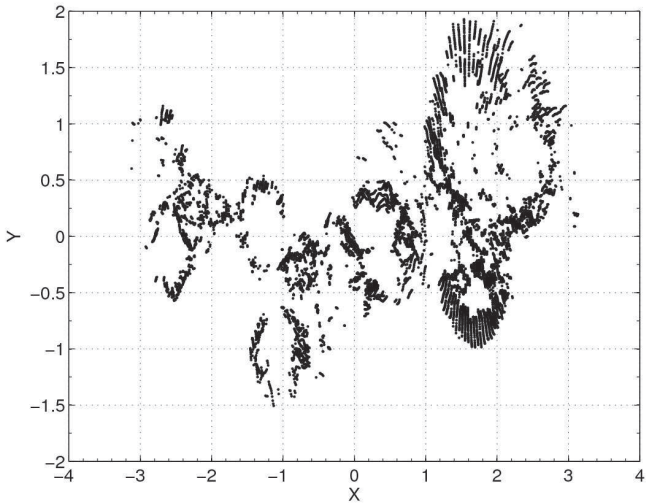}

\caption{The main shell of Cas~A's optically-emitting ejecta as
  represented in a Mercator projection. The linear scale is equal in
  all directions around any point and conformal, but the cylindrical
  map projection distorts the size and shape of large objects,
  especially towards the poles.}

\label{fig:mercator}
\end{figure}

\subsection{Kinematic Properties of the Main Shell}

The large-scale distribution of Cas~A's main shell of ejecta is shown
in Figure~\ref{fig:Rings} where knots having an expansion velocity
outside a $6000$ \kms\ sphere have been excluded. Consistent with
the recent 3D models presented in \citet{DeLaney10} based on infrared
data, we find the main shell to be dominated by morphological
structures in the form of partial or complete rings. The main
difference of our 3D reconstructions shown here to those of
\citet{DeLaney10} is increased spatial resolution (by a factor of
$\approx 4$) and more precise radial velocity data (by a factor of
$\approx 5$).

We find at least six well-defined ring-like structures on the
remnant's main shell with diameters between approximately $30\arcsec$
(0.5 pc) and $2\arcmin$ (2 pc).  In Figure~\ref{fig:mercator}, a
Mercator projection of the main shell knots is shown to illustrate the
relative scale and distribution of the rings. These rings dominate the
large-scale structure of the main shell and are present on the front,
rear, and both east and west side hemispheres.

The largest ejecta ring is located along the northern limb and is
nearly entirely redshifted (Figure \ref{fig:Rings}, top right).
Connected to it by a small bridge of ejecta knots is a much smaller
but thicker ring in the north that is entirely blueshifted (Figure
\ref{fig:Rings}, top left). These two rings contain nearly all of the
optical and infrared emission along the remnant's northern limb. A
separate but also very large, outwardly extending broken ring is
located near the base of the NE jet (Figure~\ref{fig:Rings}, bottom
left). Another ring of comparable size and radial extension lies
immediately below it.

\begin{figure*}
\centering
\includegraphics[width=0.4\linewidth]{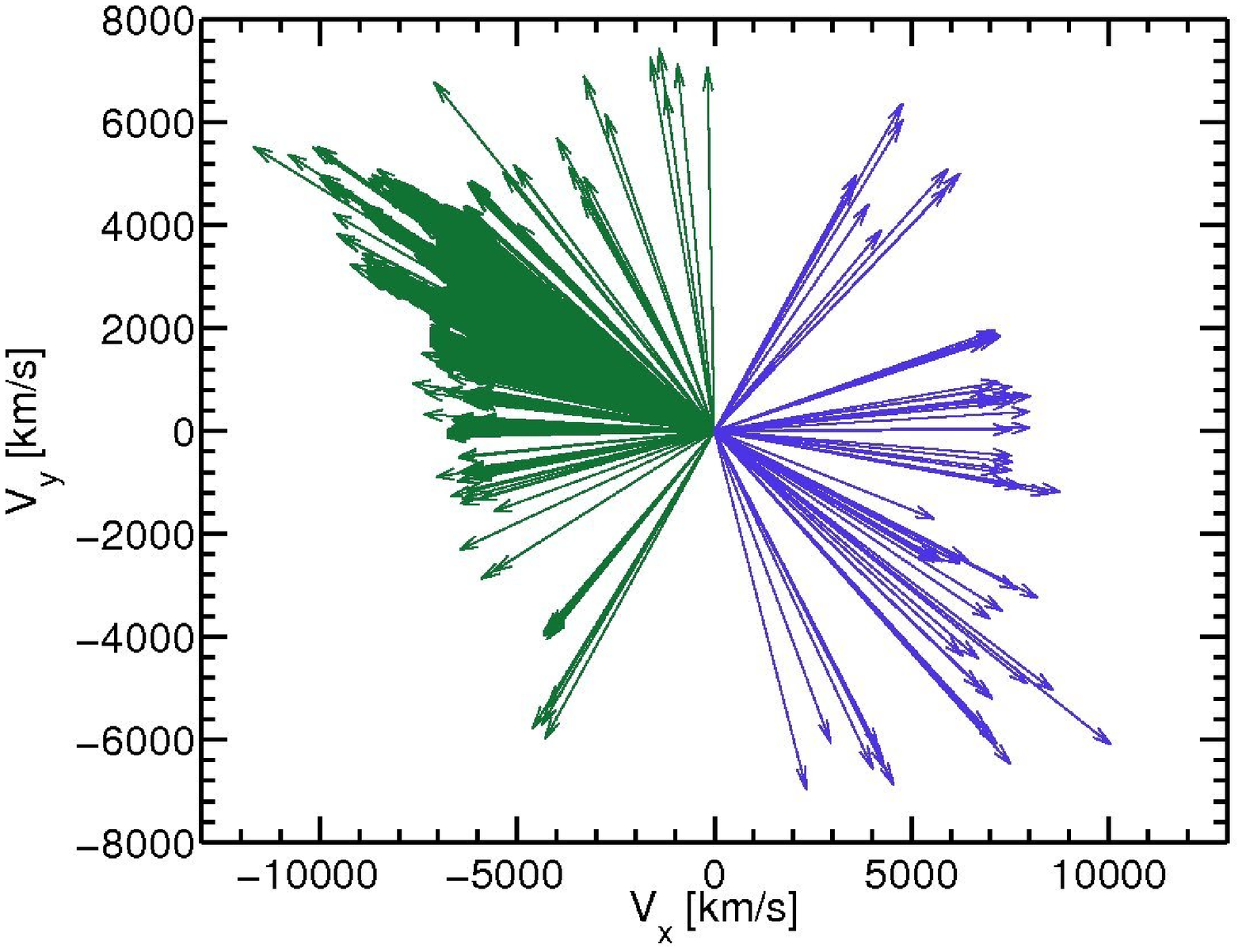}
\includegraphics[width=0.4\linewidth]{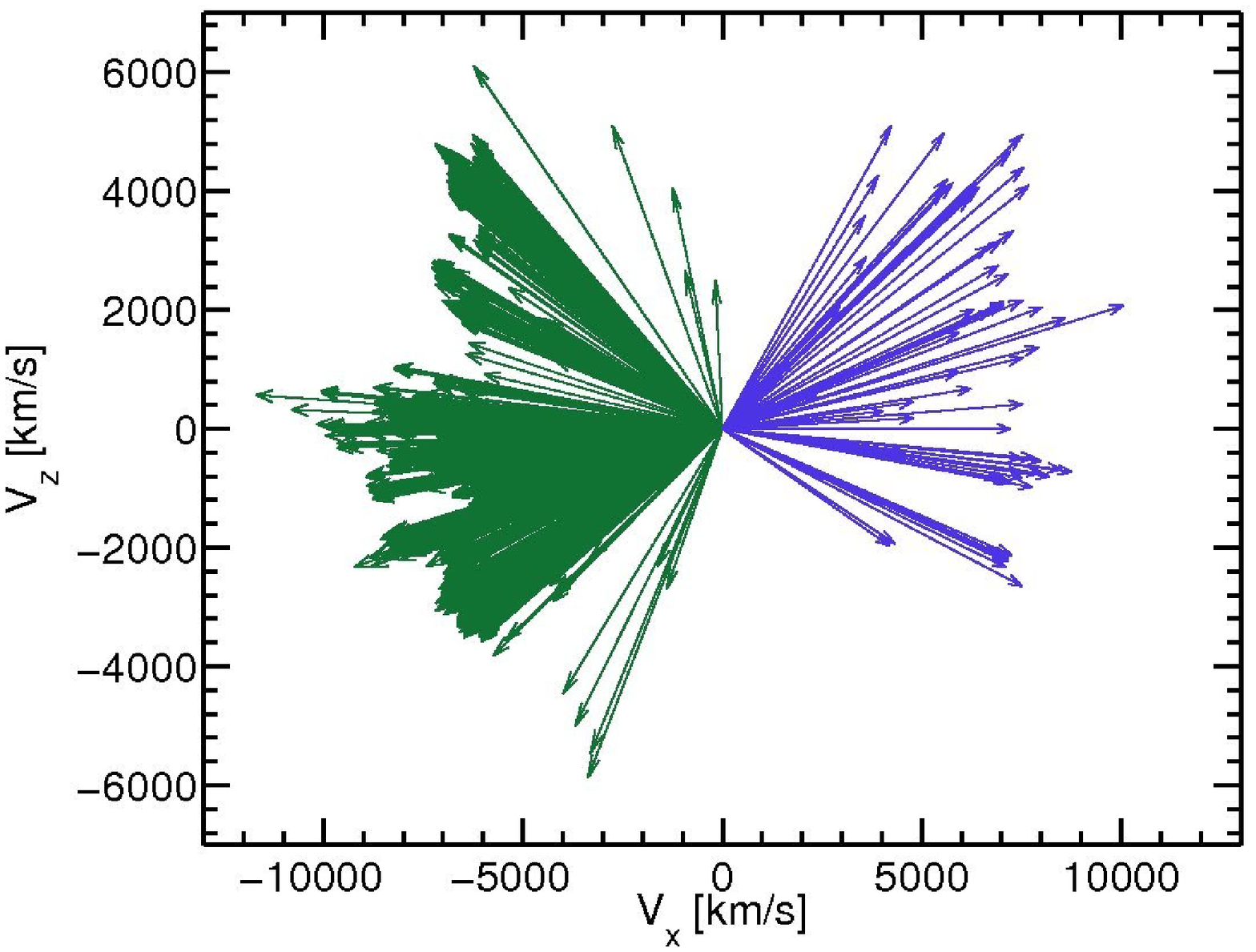}\\
\includegraphics[width=0.4\linewidth]{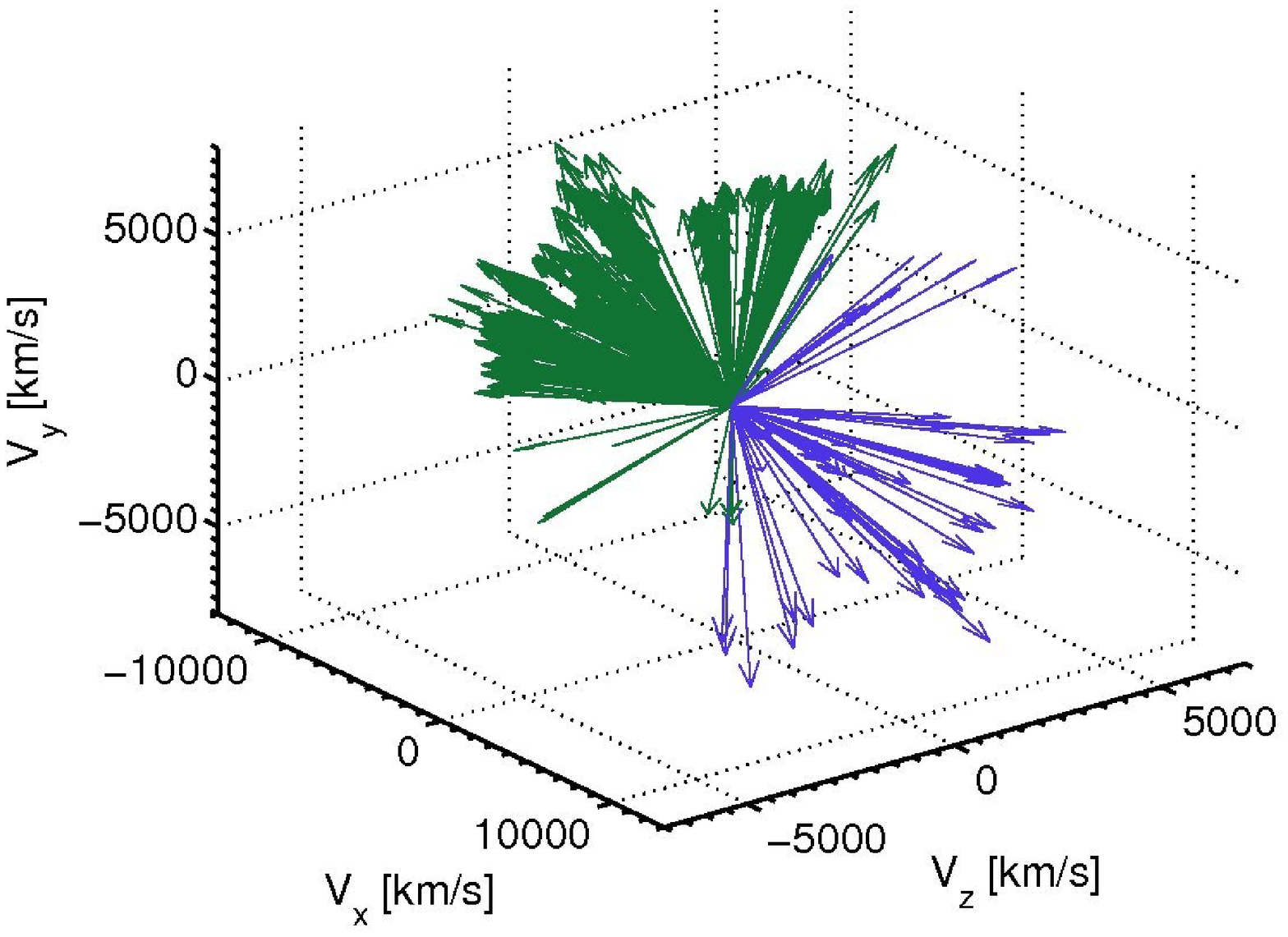}
\includegraphics[width=0.4\linewidth]{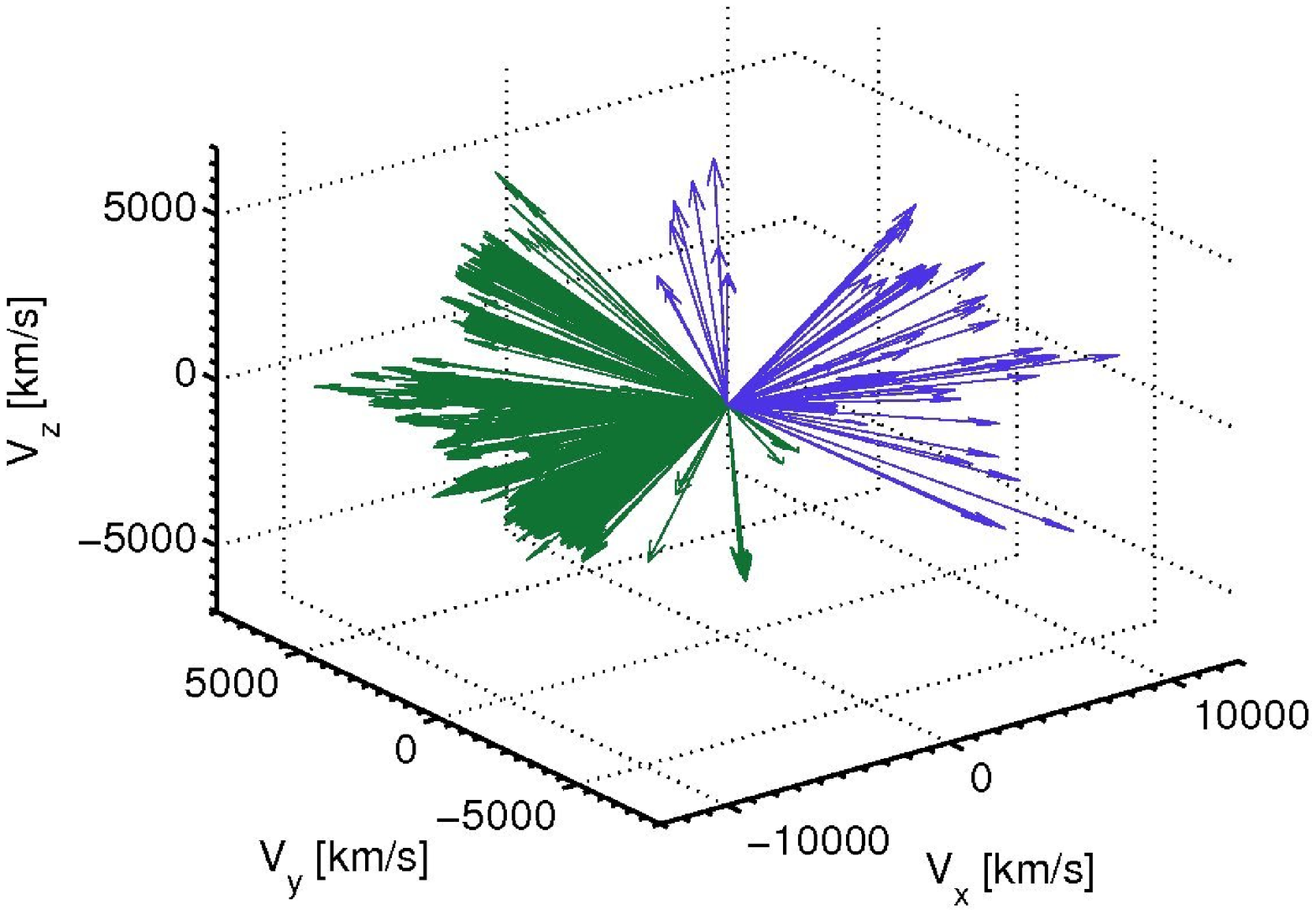}

\caption{Vector plots of all sampled outer ejecta knots. $V_x$ is
  east--west, $V_y$ is north--south, and $V_z$ is radial velocity. The
  top left panel shows the perspective as observed from Earth, with
  north up and east to the left. All knots to the east are colored
  green, and all knots to the west are colored blue. The top right
  panel shows what the jets look like as viewed along the north-south
  axis looking toward south. Two angled perspectives are shown in the
  bottom left and right panels. }

\label{fig:Jets}
\end{figure*}

Five of the six most obvious ejecta rings are actually short cylinders
giving them a crown-like appearance.  The bases of these crowns are
sometimes gently curved upward as seen in the smaller northern
blueshifted ring (Figure~\ref{fig:Rings}, top left), and the large
northern ring (Figure~\ref{fig:Rings}, top left and bottom left). The
height in velocity space of these crowns that extend radially away
from the COE is up to $1000$ \kms.  The radial extent of the broken
ring located near the base of the NE jet is largest
(Figure~\ref{fig:Rings}, top right).

Besides large-scale ejecta rings and crowned cylinders, the data show
areas with a frothy ejecta substructure down to sizes of $10\arcsec$
($\approx 5 \times 10^{17} \rm cm$). These small-scale features are
often observed at the boundaries of larger rings in tightly arranged
groupings. They can be distinguished in the Mercator projection in
Figure~\ref{fig:mercator} as the smallest rings and ellipses. Similar
small-scale ring-like structures can be seen in {\sl HST} images
(Fig.\ 5 in \citealt{Fesen01WFPC2}).

The bulk of the remnant's optically bright main shell ejecta lies
fairly close to the plane of the sky and can be contained within a
torus or thick disk structure (see Figure~\ref{fig:Rings}, bottom
left). With respect to a unit vector normal to the plane of the sky
directed from the remnant to the observer, the vector normal to the
torus' equatorial plane is tilted approximately $30\degr$ to the west
and $30\degr$ to the north, in agreement to the findings of
\citet{DeLaney10}.  We note, however, that conspicuous deviations from
a simple disk structure are observed. For example, the compact,
entirely blueshifted northern ring extends beyond the torus defined by
the bulk of the main shell. We discuss this and other deviations from
a thick disk structure more in Section~\ref{sec:explosion}.

\subsection{The NE and SW High-Velocity Streams of Ejecta}

Vector plots of outer high velocity ejecta are shown in
Figure~\ref{fig:Jets} where ejecta knots are represented by green
(east) and blue (west) vectors from the remnant's COE. An animation
has been made that shows these data rotated completely about the
north-south axis (Movie 2). Considerably more knots are detected in
the NE than in the SW. This imbalance is may be attributable in part
to higher levels of extinction in the SW compared to the NE.
Fortunately, detections in the SW jet features are numerous enough to
draw some conclusions about the relative properties of both
high-velocity ejecta features.

Previous studies of the remnant's jets lacked the kinematic resolution
and depth required to ascertain whether they form a true bipolar
structure. Our data indicate that this is indeed the case. The NE and
SW jet regions show a broad scattering of outlying, high-velocity
knots that appear to be directed in nearly opposite directions and is
suggestive of opposing flows of SN debris. Although the NE material
exhibits a slightly larger range of radial velocities compared to the
SW, we find no clear kinematic distinction between the two
distributions in terms of opening half-angle and maximum expansion
velocity.

\begin{figure*}
\centering
\includegraphics[width=0.45\linewidth]{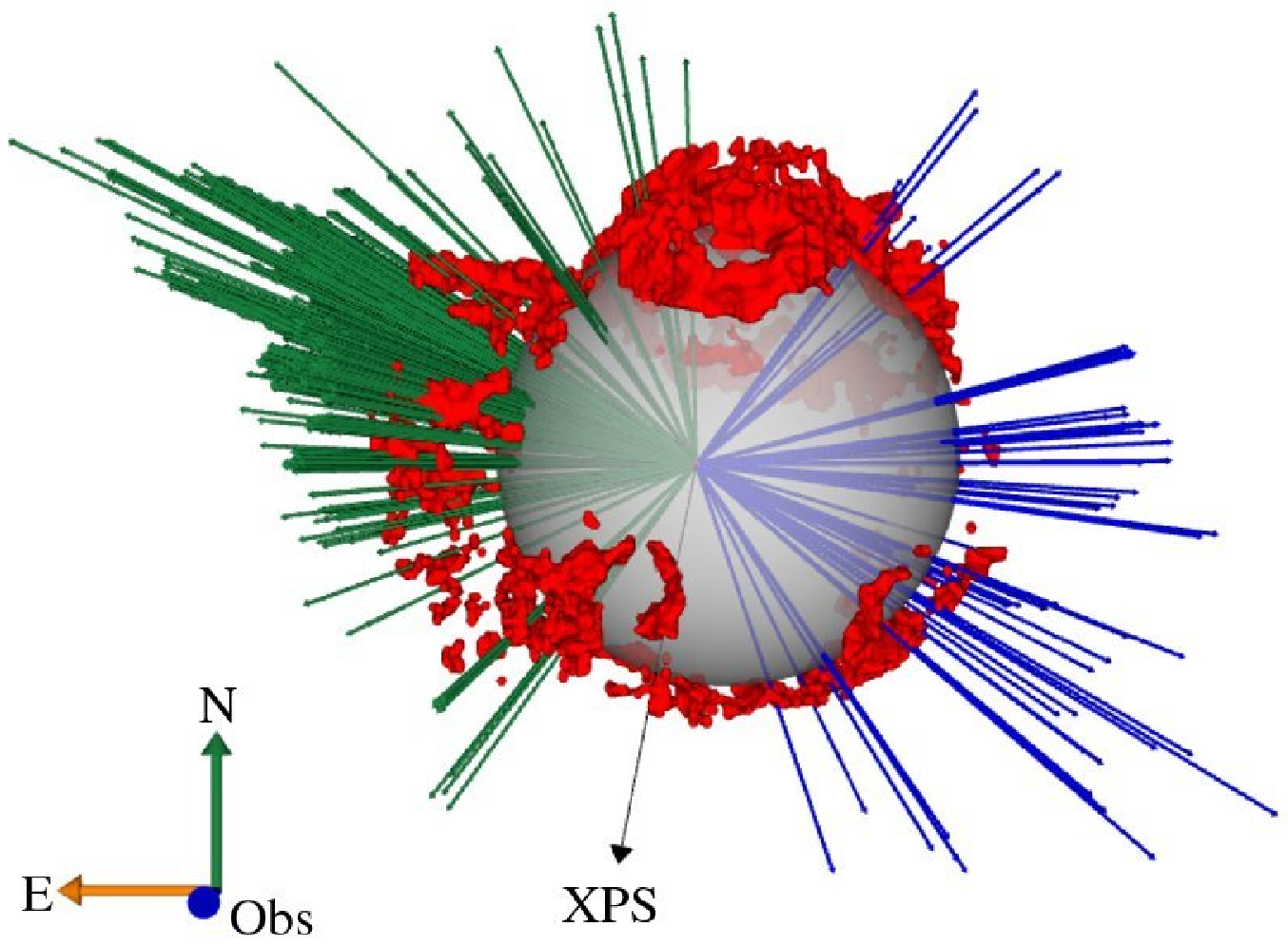}
\includegraphics[width=0.45\linewidth]{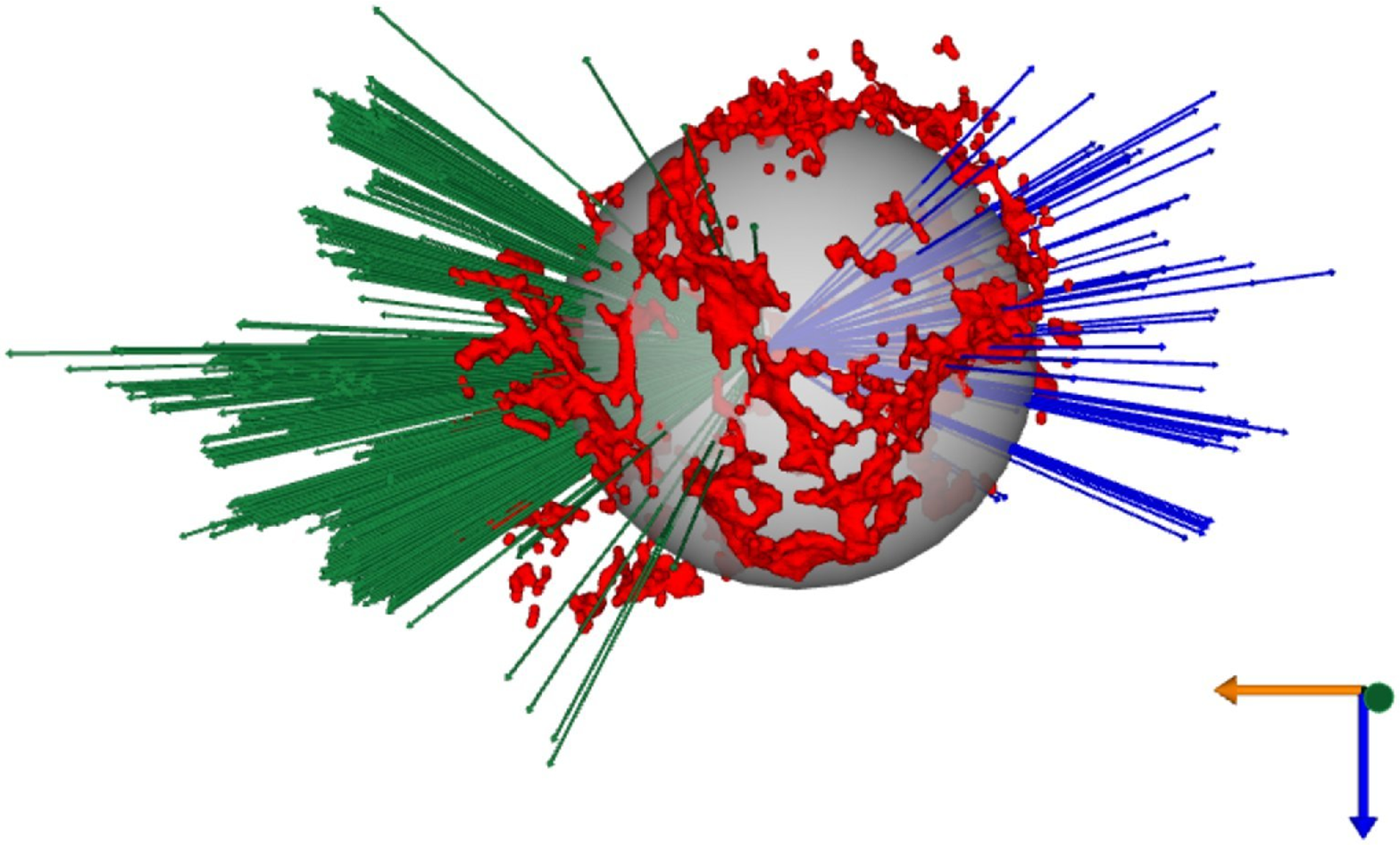}\\
\includegraphics[width=0.45\linewidth]{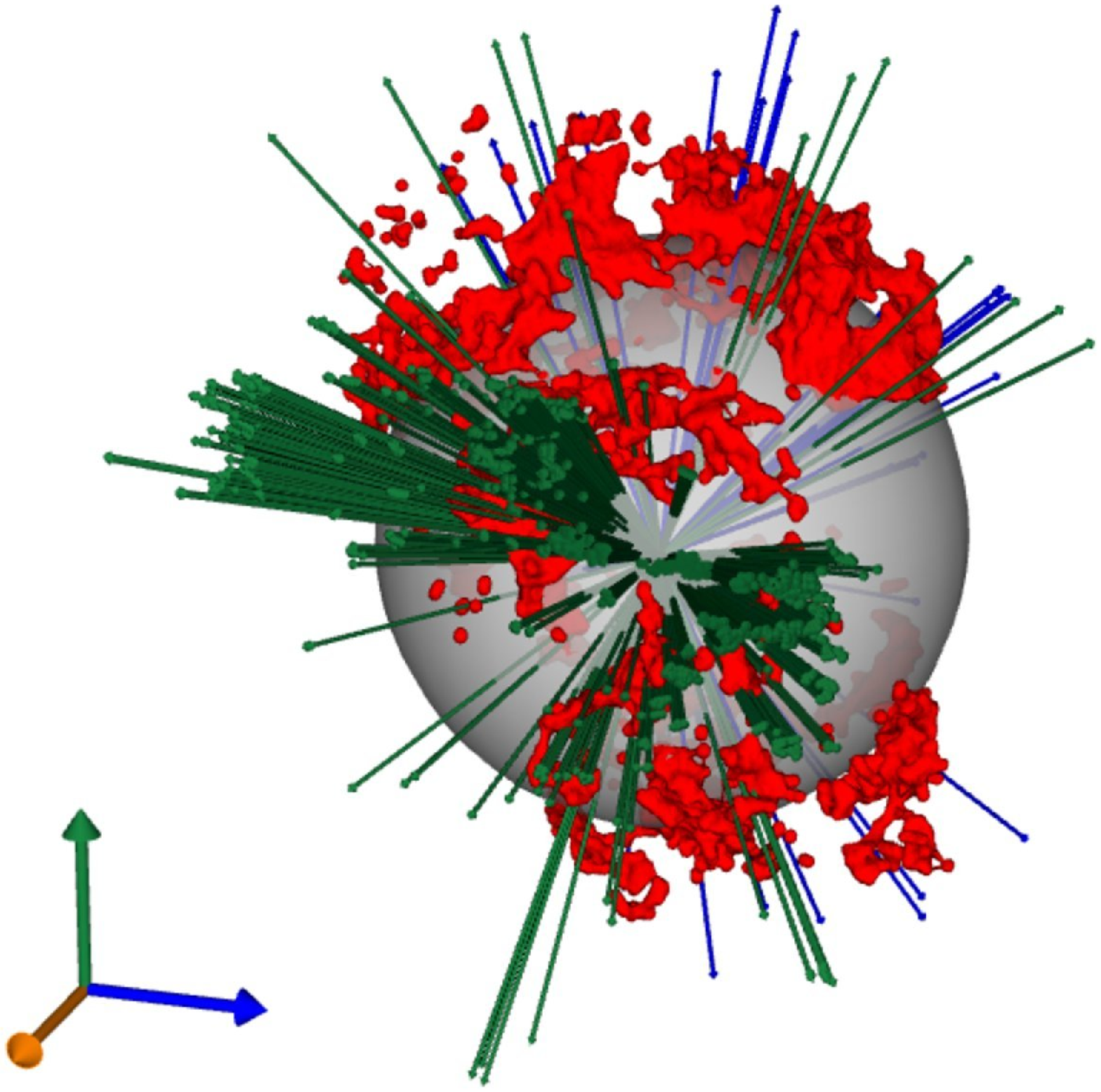}
\includegraphics[width=0.45\linewidth]{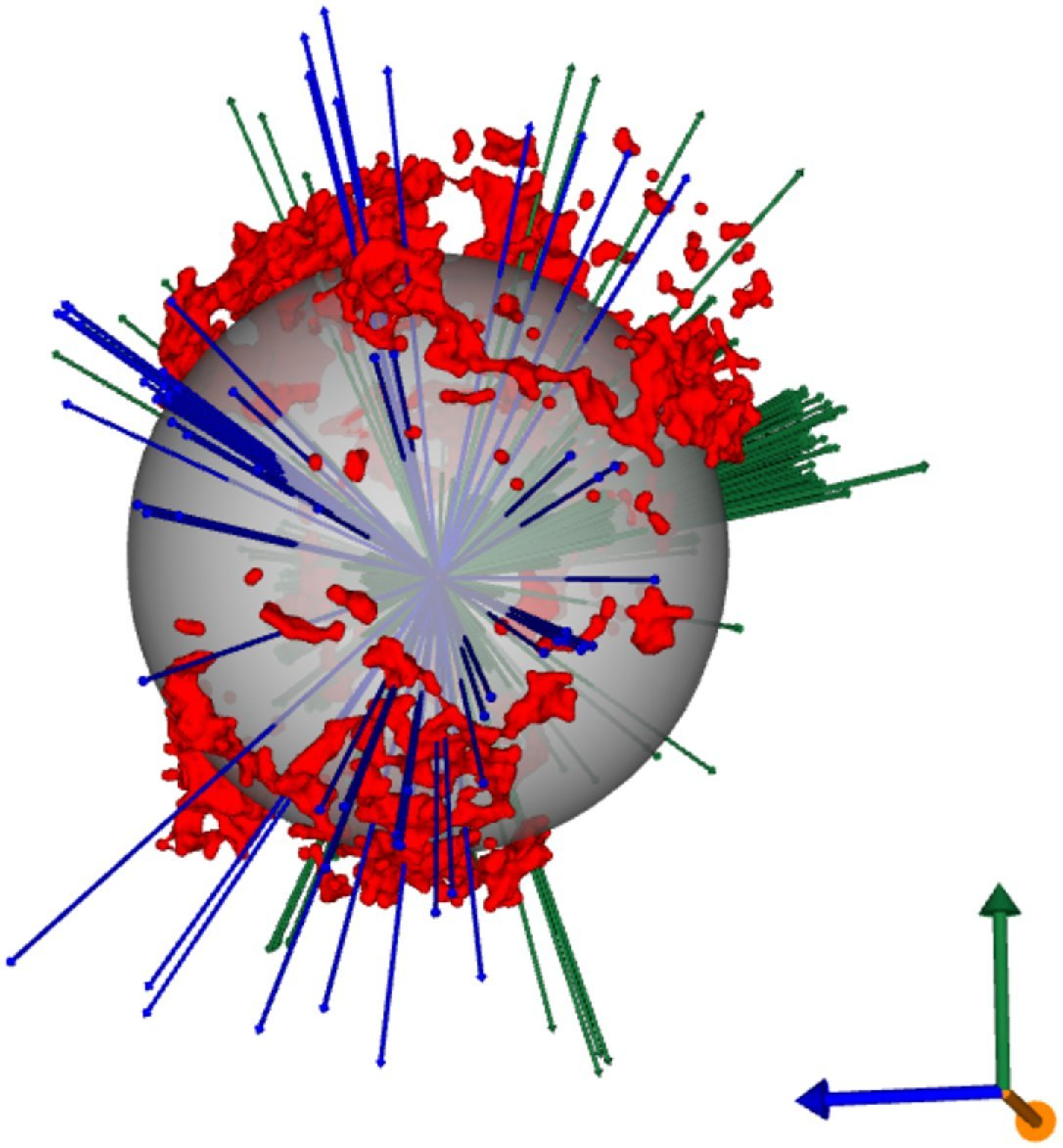}

\caption{Vector plots of all NE and SW ejecta knots (colored green and
  blue, respectively) shown together with the main shell ejecta (red).
  Four angled perspectives are shown with the upper left being as seen
  in the sky. A black arrow shows the inferred motion of the XPS with
  respect to the center of expansion \citep{Fesen06bowtie}. Refer to
  Movie 2 for an animation of these data.}

\label{fig:Jets_n_Shell}
\end{figure*}

The top left panel of Figure~\ref{fig:Jets} shows the normal observer
perspective of north up and east to the left. The distribution seen is
much like the `bowtie' plot of \citet{Fesen06bowtie} (their Fig.\ 4)
created from measured proper motions of 1825 knots seen on {\sl HST}
images taken over a 9 month period. A straight line can be drawn that
runs through the approximate center of both distributions, and this is
maintained as the perspective is changed in the various panels of
Figure~\ref{fig:Jets}.

These data reveal the brighter and richer NE jet's kinematic structure
in much greater detail than previous studies.  We find that instead of
being a few thin streams of ejecta, the NE region actually encompasses
a broad range of knot radial velocities including large redshifted
velocities up to $+5000$ \kms. Prior spectroscopic data of the NE jet
detected only a a handful of knots at these velocities
\citep{Fesen96,Fesen01Turb}.  Consequently, although the fastest knots
lie close to the plane of the sky and thus exhibit relatively modest
radial velocities, the NE jet's true structure is actually a rather
broad fan of ejecta knots with an opening half-angle between
$35-40\degr$.

In Figure~\ref{fig:Jets_n_Shell}, four perspectives of the outer
ejecta vectors with respect to the COE are shown along with the main
shell reconstruction. The top two panels are in the same orientation
as the top two panels of Figure~\ref{fig:Jets}, and use the same color
scheme to distinguish the east and west material. The possible
connection between the main shell geometry and distribution of outer
ejecta is discussed in Section~\ref{sec:jets}. Also shown in
Figure~\ref{fig:Jets_n_Shell}, is the inferred motion of the X-ray
point source (XPS) believed to be a neutron star with a carbon
atmosphere \citep{Ho09}. The direction the XPS determined originally
in \citet{Fesen06bowtie} is towards an obvious hole in the
distribution of outer ejecta and is briefly discussed in
Section~\ref{sec:XPS}.

\begin{figure*}
\centering
\includegraphics[width=0.49\linewidth]{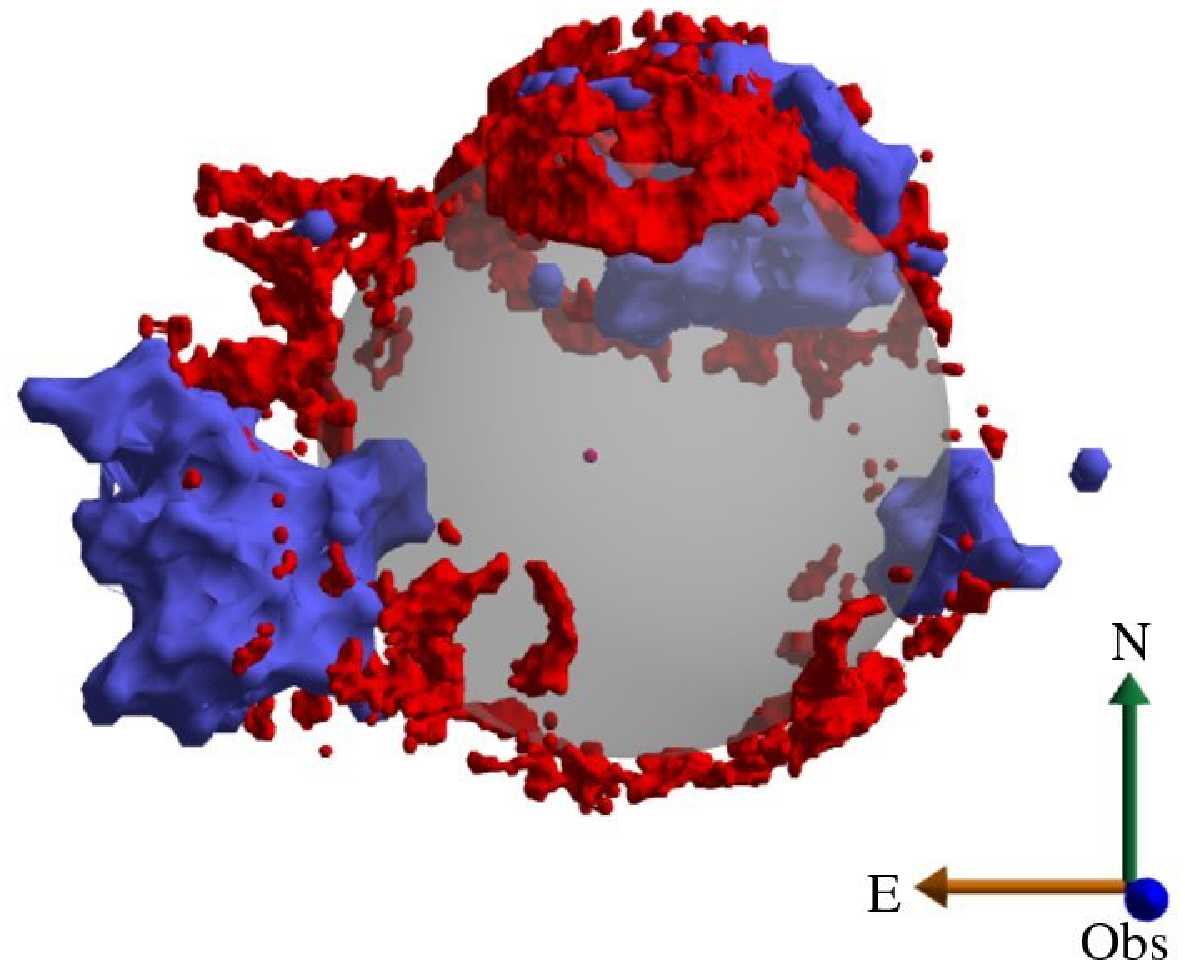}
\includegraphics[width=0.49\linewidth]{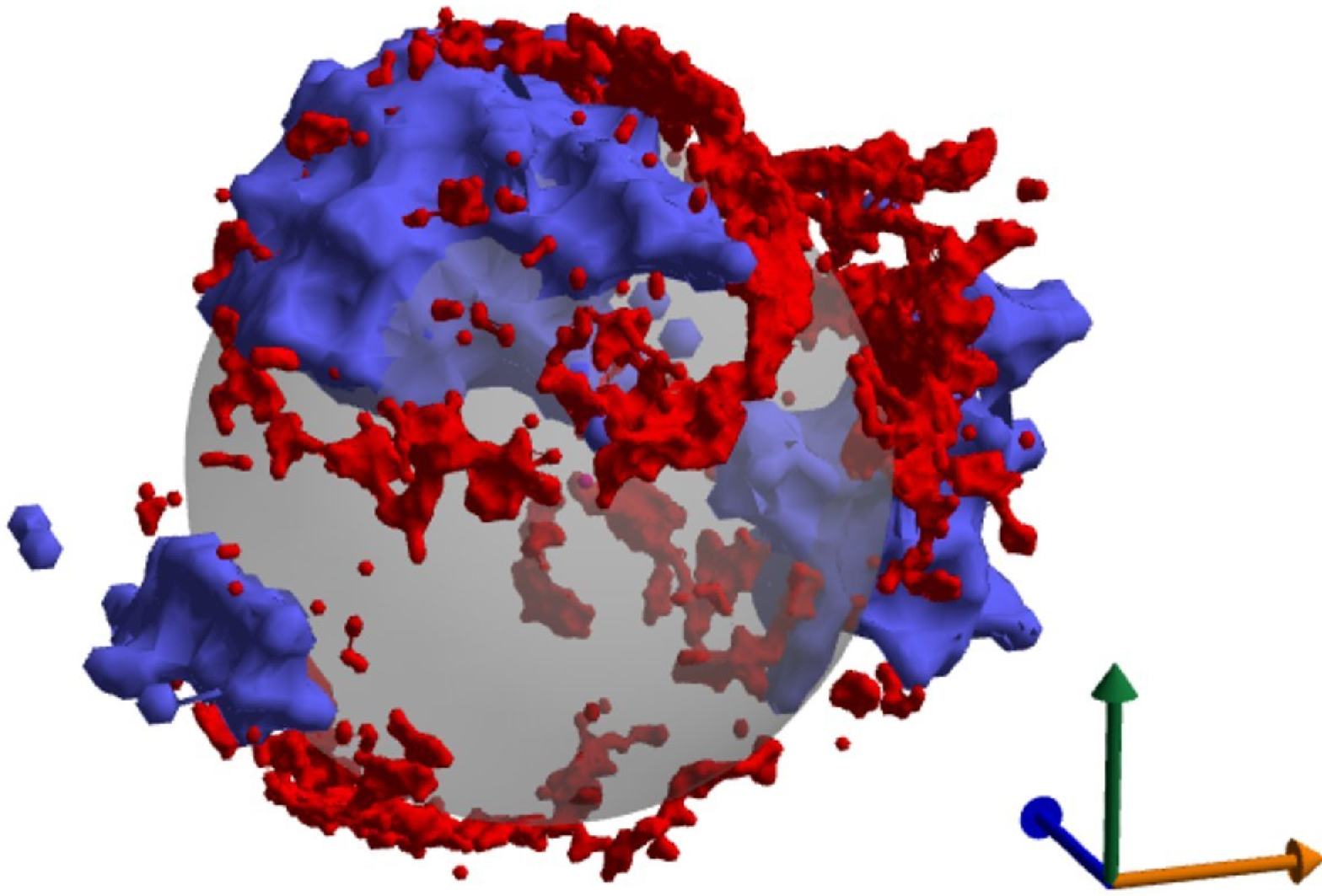}

\caption{Location of iron-rich X-ray emitting ejecta (blue) with
  respect to the main shell sulfur- and oxygen-rich optically-emitting
  ejecta (red). The X-ray data shown is from \citet{DeLaney10}. Refer
  to Movie 3 for an animation of these data.}

\label{fig:Iron}
\end{figure*}

\section{Discussion}
\label{sec:Discussion}

Although optically-emitting debris may constitute only a small
fraction of the total ejected mass in the Cas A remnant, our
observations offer superior spatial and kinematic resolution compared
to previous surveys obtained via {\sl Chandra} X-ray and {\sl Spitzer}
infrared observations. Moreover, Cas~A's fastest moving material is
best probed through optical studies. For example, optical
emission from the NE and SW jets can be traced some $90\arcsec$
farther out than in X-rays or infrared, and only a handful of outer
optical ejecta knots around the remainder of the remnant are 
detected in even the deepest X-ray images of Cas~A (see
\citealt{Fesen06bowtie}).

It is now abundantly clear from the results of our survey and previous
studies that the distribution of Cas~A's ejecta is far from being
random. At least half a dozen large and coherent ejecta ring-like
structures are observed and their size and arrangement may be
informing us about important properties of the explosion dynamics and
subsequent evolution of the expanding debris. However, interpretation
of these structures is complicated in that one must disentangle the
observed remnant properties that may originate in the explosion from
later influences related to possible post-explosion radioactive
heating, ejecta interaction with the surrounding CSM and interstellar
medium (ISM), and effects of the reverse shock on the ejecta.

An important additional caveat is that Cas~A continues to evolve and
has changed considerably in its optical appearance over the last 50
years that it has been monitored \citep{Kamper76,BK85}. On-going
propagation of the remnant's reverse shock continues to excite new ejecta,
revealing more of its entire distribution.  Hence, any conclusions
made in the present must acknowledge this evolution and recognize that
we are probably not seeing the entire remnant but only those parts
which have recently undergone shock heating.

With these limitations in mind, below we attempt to describe and
understand the nature of the observed properties of Cas~A's main shell
and outlying high-velocity ejecta.

\subsection{Ejecta Rings = $^{56}$Ni Bubbles?}
\label{sec:Iron}

\citet{Reed95} and \citet{Lawrence95} were the first to establish the
existence of conspicuous rings of reverse shocked ejecta in Cas~A.
These rings, clearly visible in the recent infrared survey of
\citet{DeLaney10}, are large (diameter $\sim$ 1 pc $\sim$ radius of
Cas A) and are most certainly structures that are related to the true
structure of the remnant's debris field and independent of any strong
influences of the remnant's CSM/ISM environment. The optical data
presented here does not alter this overall picture, but provides an
improved, more detailed examination of Cas A's expansion properties.

In Figure~\ref{fig:Iron}, we show the optically-emitting main shell
ejecta along with the location of X-ray emitting iron-rich material
measured from {\sl Chandra} data presented in \citet{DeLaney10}. An
animation showing these data sets rotated about the north-south and
east-west axes has been provided (Movie 3). Each of the three largest
concentrations of Fe-rich ejecta (i.e., along the west, north, and
southeast limbs) lie within and bounded by ring structures, strongly
suggesting a causal relationship.

\citet{DeLaney10} noted the coincidence of large ejecta rings with the
three regions of Fe--K X-ray emission and argued that these and other
less prominent features are regions where the ejecta have emerged from
the explosion as `pistons' of faster than average ejecta. In this
view, the remnant's main shell rings represent the intersection points
of these pistons with the reverse shock, similar to the bow-shock
structures described by \citet{Braun87}.

An alternative explanation first suggested by \citet{Blondin01} that
we favor is that the observed ejecta rings represent cross-sections of
large cavities in the expanding ejecta created in part by a
post-explosion input of energy from plumes of radioactive
$^{56}$Ni-rich ejecta.  \citet{Li93} have described how this input of
energy might account for the high-volume filling factor of Fe in
SN~1987A despite its small mass, and \citet{Basko94} and
\citet{Blondin01} have investigated hydrodynamic simulations based on
this model.  One possible consequence of this scenario is the
compression of surrounding non-radioactive material by the rising and
expanding bubbles of radioactive $^{56}$Ni-rich ejecta, ultimately
giving way to a ``Swiss cheese'' ejecta structure. It is important to
note that the main shell ejecta structures are observed as rings
because we are biased by the reverse shock that only excites an thin
shell of material. Thus, it is reasonable to suspect the observed
ejecta rings to be cross sections of what are actually larger
spherical geometries; i.e., bubbles.

The turbulent motions that would initiate this Ni bubble structure in
Cas\,A are not unlike recent 3D simulations of the large-scale mixing
that takes place in the shock-heated stellar layers ejected in the
explosion of a 15.5 M$_{\odot}$ blue supergiant star presented in
\citet{Hammer10}. As shown in their Figure~2, the progenitor's
metal-rich core is partially turned over with nickel-dominated fingers
overtaking oxygen-rich bullets. Although the evolution of these
simulations is strongly dependent on the internal structure of the
progenitor star \citep{Ugliano12}, it is still tempting to draw an
association between the Ni-rich outflows seen in the \citet{Hammer10}
models and the rings of Cas A.

However, there are difficulties with invoking a Ni bubble picture to
explain how the X-ray emitting Fe ejecta are framed by rings of
optical ejecta. Fe associated with the bubble effect should be
characterized by diffuse morphologies and low ionization ages
\citep{Blondin01,Hwang03}. The X-ray bright Fe we currently see,
however, is actually at an advanced ionization age relative to the
other elements and thus inconsistent with a Ni bubble origin.

It is possible that Fe associated with Ni bubbles remains
undetected. The undetected Fe would likely be of small mass compared
to what is currently observable since estimates of Cas~A's chemical
abundances are close to model predictions ($M_{\rm Fe} \sim 0.09-0.13
M_{\odot}$; \citealt{Hwang12}), and be of low density
\citep{Eriksen09}. Presently, however, there is weak evidence of
unshocked Fe ejecta in infrared data \citep{DeLaney10}. Moreover, an
exhaustive X-ray survey of Cas\,A by \citet{Hwang12} led them to
conclude that almost all of the Fe ejected by the supernova is now
well outside the reverse shock and visible in X-rays, with very little
left in the center of the remnant. Indeed, any Fe associated with the
bubble effect may be too faint and underionized to be readily
identifiable \citep{Hwang03,Hwang12}.

Thus, while considerable evidence exists that argues against
associating the X-ray bright Fe-rich material with Ni bubbles, the
large-scale and coherent ejecta rings observed in Cas~A are obviously
not random filamentary ejecta structures, and their sizes, near
spherical shapes, and ubiquity throughout the remnant are intriguingly
suggestive of a Ni bubble origin.  The strikingly tight spatial
coincidence between some optical ejecta rings and the boundaries of
Fe-rich ejecta seen in X-rays, perhaps best seen in the remnant's
large northern ring of ejecta, lends support to this picture.  Future
efforts to locate additional material internal and/or external to
Cas\,A's main shell may help clarify our understanding of the
processes responsible for the ring-shaped morphology of the majority
of Cas A's optically emitting ejecta.

Finally, we note that there is some evidence that expansion of
radioactive Ni-rich ejecta may have also influenced the ejecta
structure at smaller scales. A frothy ring-like substructure at scales
of 10$\arcsec$ ($0.2$ parsec) in high-resolution {\sl HST} images
\citep{Fesen01WFPC2} is observed around some of the large rings.
These small features may be due to fragments of Ni-rich material
pushing out surrounding material (e.g., blueshifted ring in
Figure~\ref{fig:Rings}, top left panel).  However, this substructure
should not be confused with possible influence from Rayleigh-Taylor
and Kelvin-Helmhotz instabilities that can develop in clumpy ejecta
due to their interaction with the reverse shock front.  These
processes may be the dominant ones behind the pronounced crown-like
structure of the rings (cf.\ \citealt{Basko94,Blondin01}).

\subsection{Main Shell Geometry and Velocity Asymmetry}
\label{sec:explosion}

In addition to the large-scale ring structures of the main shell,
there are two other observed properties of the main shell ejecta that
may be related to the explosion dynamics. The first is the $-4000$ to
$+6000$ \kms\ radial velocity asymmetry. This asymmetry has long been
recognized in even the earliest optical observations of the remnant
made by \citet{Minkowski68}, and later more fully appreciated by
observations of \citet{Lawrence95} and \citet{Reed95}. The second is
the torus-like geometry observed for the bulk of the remnant's optical
and infrared bright ejecta, discussed at length by \citet{Markert83}
and more recently by \citet{DeLaney10}.

Interpreting these properties is complicated given the problem of
distinguishing whether the remnant's observed geometry is a faithful
representation of the explosion dynamics or has been significantly
influenced by an inhomogeneous local CSM and ISM. \citet{Reed95}
argued that environmental effects were significant in understanding
the remnant's blue and redshift velocity asymmetry. They concluded
that the density of the surrounding medium is greater on the
blueshifted near side, which would inhibit forward expansion and
result in an apparently redshifted COE. In support of this conclusion,
they cite the observed asymmetrical distribution of radial velocities
of the circumstellar QSF knots, a majority of which (76\%) are
blueshifted.

However, \citet{DeLaney10} and \citet{Isensee10} note that there are
major structural differences between the front and back interior
surfaces of Cas~A.  These differences in brightness and structure
could be due to different masses, densities, and energies in the two
different directions suggesting the observed velocity asymmetry might
be intrinsic to the explosion itself.

\citet{DeLaney10} further argue that the torus-like distribution of
bright main shell ejecta is a result of the explosion dynamics. They
conclude that although interaction with the CSM affects the detailed
appearance of the remnant, the bulk of the symmetries and asymmetries
in Cas~A are intrinsic to the explosion. In their framework, the
remnant is the result of a flattened explosion where the highest
velocity ejecta were expelled in a thick torus tilted not far off the
plane of sky in a number of large-scale pistons.

In possible support of their picture, \citet{DeLaney10} noted that
\citet{Fesen01Turb} found that relatively small angles off the sky
plane (i.e., $<30\deg$) seem to be the rule for the remnant's small,
fastest-moving outlying ejecta knots. That is, although many hundreds
of outlying ejecta knots can be seen around most of Cas~A's periphery,
none have been detected close to its center.

\begin{figure*}
\centering
\includegraphics[width=0.48\linewidth]{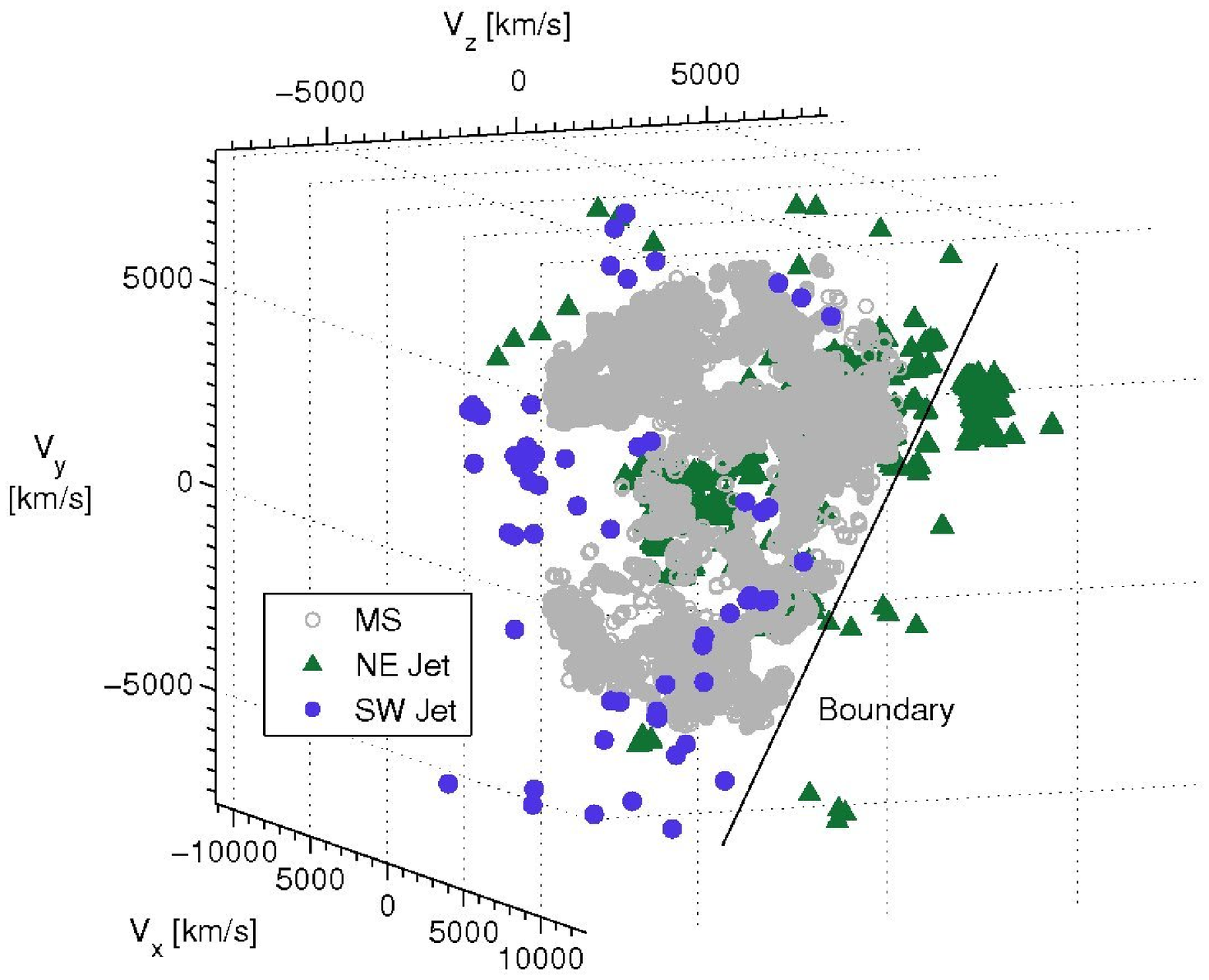}
\includegraphics[width=0.45\linewidth]{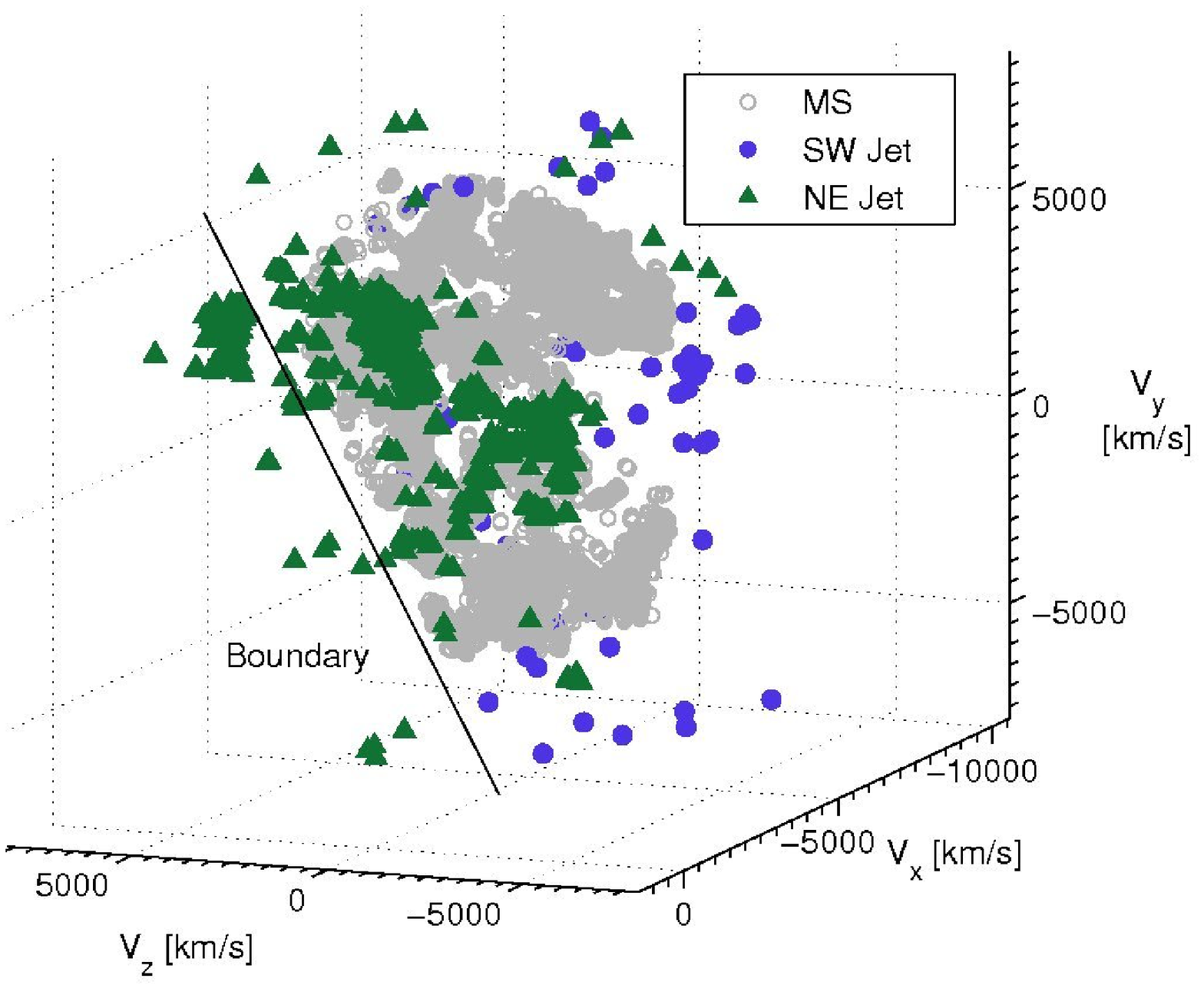}

\caption{Side-looking perspectives of Cas A showing non-uniform
  distribution of material. The SW material (blue) and main shell
  ejecta (gray) cut off at a boundary defined by an angled plane. The
  NE material (green) does not cut off at the boundary, and this may
  due to interaction with nearby ISM/CSM H-rich material seen in
  H$\alpha$.}

\label{fig:Boundary}
\end{figure*}

Our extensive survey also was unable to find any small, very
high-velocity knots within the main shell and projected near the
remnant's center.  Rather than supporting a torus-like ejecta
arrangement, however, this could indicate that a near tangent viewing
angle may be important for a knot's optical
visibility. \citet{Fesen01Turb} originally proposed that a near
tangent viewing factor might hold for both the remnant's outer knots
and its main shell.  If true, this would mean some portion of the
remnant's facing and rear sides may remain undetected, and thus the
currently observed torus-like configuration of bright emission not
reflective of the remnant's true 3D structure.

The notion of a disk geometry for the Cas A remnant that happens to
lie fairly near to the plane of the sky is also inconsistent with  
recent infrared observations that detected some of the remnant's
interior, unshocked ejecta. \citet{Isensee10} observed redshifted
[\ion{Si}{3}] and [\ion{O}{4}] line emissions from interior debris
along a sight line close to remnant center exhibiting velocities
approaching 5000 km s$^{-1}$. These velocities are comparable to the $
-4000$ to $+6000$ km s$^{-1}$ maximum radial velocities seen for main
shell ejecta, as well as the inferred maximum transverse velocity of
around 6300 km s$^{-1}$ derived from proper motions ($\approx
0\farcs39$ yr$^{-1}$) of main shell ejecta knots assuming a remnant
distance of 3.4 kpc \citep{vandenBergh83,Thorstensen01}.

Further support of this viewpoint are some conspicuous departures from
a thick-disk geometry in the distribution of main shell
ejecta. Rear-facing ejecta seem to cut off abruptly along an angled
plane, while the front-facing ejecta do not. In particular, the
northern, blueshifted ring forms a conspicuous bulge from the apparent
torus (see bottom left panel of Figure \ref{fig:Rings}).  There are
aspects of the asymmetric distribution of the remnant's outermost
ejecta that also indicate possible influence of an inhomogeneous
environment. In Figure~\ref{fig:Boundary}, the distribution of
material as observed from two sides of Cas~A looking toward the base
of the NE and SW jets is shown. The material in the SW jet region
largely follows the abrupt boundary cutoff, as does the main shell
ejecta. Ejecta in the the NE jet region, however, do not follow this
cut-off. Deep H$\alpha$ images of Cas~A show evidence for much more
H-rich CSM/ISM material in the eastern limb than in the western limb
\citep{Reynoso97,Fesen01Turb}, which is consistent with the cut-off
being due to environmental effects.

\subsection{Opposing High-Velocity Ejecta Outflows}
\label{sec:jets}

The nature of the anomalously high velocities of ejecta in the NE and
SW regions has long been a puzzle.  \citet{Minkowski68} suggested the
NE jet might be the sole surviving part of an outer, high-velocity
shell that has been subsequently decelerated in all other directions.
Alternatively, because the distribution of emitting gas may not
necessarily be the same as the actual distribution of gas in Cas~A,
the jets may simply be the most visible part of a larger population of
the remnant's outer, high velocity knots \citep{Fesen96}.

Along similar lines, asymmetric debris structures produced by
circumstellar interaction has been theoretically modeled by
\citet{Blondin96} who show how a jet-like feature of SN ejecta can be
generated in the progenitor's equatorial plane from pole/equator
density gradients in the local CSM.  Thus, these high-velocity regions
may be secondary features caused by instability-powered flows from an
equatorial torus, where the explosion axis is loosely defined by the
X-ray iron-rich regions found in the southeast and northwest
\citep{Burrows05,Wheeler08,Rest11}.

However, the very prominent rupture-like features in the main shell
near the base of the NE jet visible in X-ray, optical, IR, and radio
images (cf.  \citealt{Hughes00,BK85,Ennis06,Anderson95,DeLaney10}) are
certainly suggestive of an explosive formation process. Although
perhaps not indicative of a jet-induced explosion, there is
nonetheless substantial evidence that the NE and SW jets are somehow
associated with core-collapse explosion dynamics in some way.

For example, analyses of {\it Chandra} X-ray spectra indicate that the
NE jet could not be formed through interaction with a cavity or lower
density region in the CSM \citep{Laming03,Laming06}. The remnant's
ejecta would have expanded rapidly into a CSM cavity, and led to a
density too low for electron-ion equilibration to match the observed
values. \citet{Laming06} estimate that a factor of two more explosion
energy went into the NE jet direction compared to other regions.

The location of chemically distinct optically-emitting knots in both
jets is also consistent with an unusual, high-velocity ejection of
underlying mantle material. Knots exhibiting a mix of H$\alpha$,
[\ion{N}{2}], [\ion{O}{2}], [\ion{S}{2}], and [\ion{Ar}{3}] line
emissions are {\it only} observed in the jet regions, suggesting an
eruptive and turbulent mixing of the underlying S, O, and Ar rich
material with photospheric H- and N-rich layers
\citep{Fesen91,Fesen96,Fesen01Turb}.  Furthermore, optical knots in
the NE and SW jets lying farthest out and possessing the highest
ejection velocities show no detectable emission lines other than those
of [\ion{S}{2}], suggesting an origin from the S-Si-Ca-Ar rich layer
deep inside the progenitor star
\citep{vandenBergh83,Fesen96,Fesen01Turb,Fesen01WFPC2}.

As shown in Figure~\ref{fig:Jets}, we find that the overall conical
distributions of the NE and SW jet regions exhibit comparable maximum
expansion velocities and are broadly anti-parallel in an orientation
consistent with a bipolar jet-counterjet structure. However, the
observed opening half-angles of these two flows ($\sim 40\degr$) is
quite broad and is not what would be anticipated in a
highly-collimated (opening half-angle $< 10\degr$) jet-induced
explosion. Thus, the broadness and estimated low energies of the NE
and SW jets make it unlikely that they are signatures of a jet-induced
explosion scenario (e.g.,
\citealt{Khokhlov99,Wheeler02,Akiyama03,Fryer04,Laming06}).

There appears to be only a tenuous connection between the NE and SW
jet regions and any of the main shell's ejecta rings. The centers of
cones broadly defining the NE and SW ejecta streams as sampled in our
survey only loosely point to centers of ejecta rings. The respective
distributions are illustrated in Figure~\ref{fig:Jets_n_Shell}, where
the outer knot vectors are plotted with the main shell surface
reconstruction. Particularly suggestive is the NE jet region, where
the bulk of high-velocity material appears to be emanating from a
large main shell ring that has a pronounced radial extension. In this
regard, the main shell rings and the jet-like streams of higher
velocity debris could be dynamically related.

However, for both the NE and SW jets, many knots have trajectories
that do not pass through any main shell rings. In light of the large
number of faint, outermost high-velocity jet knots, it is possible
that additional material at even higher positive and negative radial
velocities may lie undetected, broadening the jets even more.  Future
work mapping the chemical abundances of jet knots to their kinematics
might contribute to a better understanding of their nature and
possible relationship to the main ejecta shell.

\subsection{Motion of X-ray Point Source}
\label{sec:XPS}

An additional clue to the nature of the explosion dynamics may come
from understanding the properties of the inferred motion of the
central XPS. The location of the XPS $\approx 7\arcsec$ to the
southeast of the estimated COE (P.A.$\approx 170\degr$) indicates a
transverse velocity `kick' of $\approx 350$ \kms\ imparted to the
compact object during the explosion \citep{Fesen06bowtie}.

Intriguingly, the projected line connecting the NE and SW jets lies
nearly perpendicular to the inferred direction of the XPS (see
Figure~\ref{fig:Jets}). Assuming the NE/SW axis to be the most
significant in the original core-collapse, this runs counter to most
jet-induced explosion models that predict that the neutron star will
undergo a kick roughly aligned with the jet axis
\citep{Burrows96,Fryer04}.

There is a noticeable absence of outlying ejecta along the axis of the
XPS's inferred motion \citep{Fesen01Turb,Fesen06bowtie}.  With the
possible exception of unlikely and extremely localized extinction
conditions, deep imaging of Cas~A surveying the entire remnant ensures
that there is no observational bias that could explain the lack of
detection of ejecta in these regions. Although possibly just a
coincidence, it is interesting that the XPS should be moving with a
trajectory in line with the only two significant gaps of the outer
ejecta which are located nearly perpendicular to the NE-SW jet axis.

We note that \citet{Smith09} showed that strong [\ion{Ne}{2}] emission
is observed symmetrically around the XPS, with an axis also close to the
direction of the inferred kick.  Although \citet{Smith09} determined
that there is not enough mass/energy in the Ne-crescents to affect the
XPS motion, they may perhaps arise due to the same dynamical
asymmetries that led to the XPS motion \citep{DeLaney10}.

\subsection{Comparison to Other Young SNRs}

Recent 3D reconstructions of other young CCSN remnants have suggested
that many of the kinematic properties observed in Cas~A are not
unique. Although only a handful of known SNRs are appropriate for this
kind of analysis, those that have been studied have revealed evidence
of the same large-scale ejecta rings, velocity asymmetries, and
high-velocity bipolar asymmetries as observed in Cas~A.

For example,\citet{Vogt10} mapped the [\ion{O}{3}] $\lambda$5007 line
emission dynamics of the young oxygen-rich supernova remnant 1E
0102.2-7219 in the Small Magellanic Cloud. They confirmed earlier
Doppler measurements of \citet{Eriksen01} that showed the presence of
large ejecta rings like those observed in Cas~A. Also like in Cas~A,
they found 1E 0102.2-7219 to have an overall velocity asymmetry in its
bulk material, such that redshifted material contained fainter clumps
but with higher velocities than the blueshifted material.

A similar [\ion{O}{3}] $\lambda$5007 survey of the young oxygen-rich
supernova remnant N132D in the Large Magellanic Cloud by
\citet{Vogt11} also uncovered properties similar to Cas~A. They found
the majority of the ejecta to form a ring of $\sim$12 pc in diameter
inclined at an angle of $\sim 25\degr$ to the line of sight, and
evidence of a polar jet associated with a very fast oxygen-rich
knot. The survey led \citet{Vogt11} to speculate that the overall
observed shape of the SNR to have been strongly influenced by the
pre-supernova mass loss from the progenitor star.

Work by \citet{Winkler09} on the core-collapse SNR G292.0+1.8 shows
that it too has properties like those seen in Cas~A. Proper motion
measurements of G292.0+1.8's fast filaments exhibit systematic motions
outward from a point near the center of the radio/X-ray shell that
lies $46\arcsec$ northwest from the young pulsar PSR J1124-5916. The
inferred motion of the pulsar has a transverse velocity of 440 \kms,
which is close to that observed in Cas~A's XPS. Additionally, the
fastest ejecta in G292.0+1.8 form a bipolar outflow along an axis
oriented roughly north--south in the plane of the sky.  The remnant
appears to have undergone a complex evolution resulting from a
possibly asymmetric SN explosion and interaction with non-uniform
ambient ISM (see also \citealt{Braun86,Park02}), which is a scenario
not too different from the one we envision for Cas~A.

\subsection{Connections to Extragalactic SN Observations}

Finally, we consider how the kinematic properties of Cas~A discussed
here may be linked to phenomena observed in the late-time optical
emissions of extragalactic supernovae.  \citet{Milisavljevic12} summed
all main shell spectra presented in Section \ref{sec:Observations}
into a single, integrated spectrum mimicking what the remnant would
appear as as an unresolved extragalactic source.  Similarities were
seen between this integrated Cas~A spectrum and several late-time
optical spectra of decades-old extragalactic SNe.

Particularly well-matched with Cas~A were SN~1979C, SN~1993J,
SN~1980K, and the ultra-luminous supernova remnant in NGC~4449.
\citet{Milisavljevic12} found that both Cas~A and decades old
extragalactic SNe show pronounced blueshifted emission with
conspicuous line substructure in [\ion{O}{1}], [\ion{O}{2}],
[\ion{O}{3}], [\ion{S}{2}], and [\ion{Ar}{3}].  Since the emission
line substructure observed in the forbidden oxygen emission line
profiles of Cas~A are associated with the large-scale rings of ejecta,
we suggest that identical features in the intermediate-aged SNe which
have often been interpreted as ejecta `clumps' or `blobs' are, in
fact, probable signs that large-scale rings of ejecta are common in
SNe.  The link is strongest between Cas~A and SN~1993J, given that
these were both Type IIb explosions and that they exhibit extremely
similar forbidden oxygen profiles despite the three centuries of
evolution that separate them.

\section{Conclusions}
\label{sec:Conclusions}

We have presented 3D kinematic reconstructions of the optical emission
from the supernova remnant Cas~A based on radial velocity measurements
extracted from long-slit and multi-slit spectra. This data set is of
high spatial and kinematic resolution and encompasses the NE and SW
streams of high velocity ejecta jets that until now have never been
surveyed to this depth.

The major results and conclusions of this study are as follows:

1) We confirm the findings of several previous kinematic studies of
Cas~A that show the bulk of the remnant's main shell ejecta to be
arranged in several well-defined complete or broken ring-like
structures. These ring structures have diameters that can be
comparable to the radius of the remnant ($\sim$1 pc). Some rings show
considerable radial extensions giving them a crown-like appearance,
while other rings exhibit a frothy, ring-like sub-structure on scales
of $\sim$ $0.2$ pc.

Such large-scale ejecta rings may be a common phenomenon of young,
core-collapse supernova remnants. Evidence for this comes in part from
lumpy emission line profile substructure seen in both an integrated
Cas~A spectrum and several late-time optical spectra of extragalactic
supernovae obtained years to decades after explosion. Because this
line substructure can be directly mapped to the large-scale rings of
our Cas~A reconstruction, we suggest a similar origin may apply to
unresolved extragalactic remnants.

2) The bulk of Cas~A's optically bright ejecta populate a torus-like
geometry that is tilted approximately 30$\degr$ with respect to the
plane of the sky and exhibit a $-4000$ to $+6000$ \kms\ radial
velocity asymmetry. Unlike the conclusion reached by
\citet{DeLaney10}, we suggest that this observed geometry and velocity
asymmetry is not representative of the true overall kinematic
properties of the original explosion. Instead, an observational bias
caused by a near tangent viewing angle effect and interaction with an
inhomogeneous CSM/ISM environment has likely contributed to some of
Cas~A's presently observed kinematic properties.

3) The size, shape, and ubiquity of large-scale, reverse shock heated
ejecta rings in Cas~A are consistent with a bubble-like interior
structure. It is possible that some of these structures were generated
by the input of energy from radioactive $^{56}$Ni rich ejecta that
produced low density bubbles surrounded by non-radioactive,
intermediate mass element-rich ejecta. While the spatial coincidence
between X-ray emitting Fe-rich material and some of the remnant's
large ejecta rings is suggestive of such a origin, the high ionization
age of the X-ray bright Fe suggests that this material is not directly
associated with Ni bubbles.

4) Our deep optical survey shows the remnant's NE and SW jet features,
which contain Cas~A's highest velocity ejecta, to be unexpectedly
broad streams of ejecta knots with comparable conical opening
half-angles of approximately $40\degr$.  The jets appear to lie in an
orientation consistent with an opposing and wide bipolar outflow.  The
broadness together with low energy estimates of the NE and SW jets
argue against them being associated with a jet-induced
explosion. However, several kinematic and chemical properties support
the view that they were formed during core collapse and independent of
effects from the local CSM or ISM.  Faint ejecta in both jets at even
higher expansion velocities may lie as yet undetected.

We end by noting that similar radial velocity data sets spanning
X-ray, optical, and infrared wavelengths of other young SN remnants
may provide additional insights and tests of various core-collapse SN
models.  A comparison of Cas~A's properties with those of other young
remnants, especially ones supplemented with chemical abundance
analyses in addition to kinematics, may also help to further constrain
details of the explosion mechanism, post-explosion dynamics, ejecta
instabilities, and explosive nucleosynthesis.  With the growing
success obtaining multi-perspective optical spectra of light echoes
associated with other historical Galactic supernovae e.g., SN\,1572
\citep{Krause08b}, and those nearby in the LMC and SMC \citep{Rest05},
the possibility of developing 3D reconstructions of other young CCSN
supernova remnants with known SN subtypes may be possible in the
future.

\section{Acknowledgments}

We thank an anonymous referee for comments that improved this paper's
content and presentation. D.\ Patnaude, R.\ Chevalier, J.\ Thorstensen
and G.\ Wegner provided helpful discussions and comments.  T.\ DeLaney
kindly provided infrared and X-ray data in advance of
publication. David Adalsteinsson provided considerable support with
the DataTank software (http://www.visualdatatools.com), which was used
to prepare some figures and the animations. Surface reconstruction was
aided with the use of MeshLab (http://meshlab.sourceforge.net), a tool
developed with the support of the 3D-CoForm project. This material is
based upon work supported by the National Science Foundation under
Grant No.\ AST-0908237.


\begin{thebibliography}{68}

\bibitem[{{Akiyama} {et~al.}(2003){Akiyama}, {Wheeler}, {Meier}, \&
  {Lichtenstadt}}]{Akiyama03}
{Akiyama}, S., {Wheeler}, J.~C., {Meier}, D.~L., \& {Lichtenstadt}, I. 2003,
  \href{http://dx.doi.org/10.1086/344135}{\apj, 584, 954}

\bibitem[Anderson \& Rudnick(1995)]{Anderson95} Anderson, M.~C., \&
Rudnick, L.\ 1995, \apj, 441, 307

\bibitem[{{Basko}(1994)}]{Basko94}
{Basko}, M. 1994, \href{http://dx.doi.org/10.1086/173983}{\apj, 425, 264}

\bibitem[Besel \& Krause(2012)]{Besel12} Besel, M.-A., \& Krause, O.\
2012, \aap, 541, L3

\bibitem[{{Blondin} {et~al.}(2001){Blondin}, {Borkowski}, \&
  {Reynolds}}]{Blondin01}
{Blondin}, J.~M., {Borkowski}, K.~J., \& {Reynolds}, S.~P. 2001,
  \href{http://dx.doi.org/10.1086/321674}{\apj, 557, 782}

\bibitem[{{Blondin} {et~al.}(1996){Blondin}, {Lundqvist}, \&
  {Chevalier}}]{Blondin96}
{Blondin}, J.~M., {Lundqvist}, P., \& {Chevalier}, R.~A. 1996,
  \href{http://dx.doi.org/10.1086/178060}{\apj, 472, 257}

\bibitem[Blondin et al.(2003)]{Blondin03} Blondin, J.~M., 
Mezzacappa, A., \& DeMarino, C.\ 2003, \apj, 584, 971 

\bibitem[{{Braun} {et~al.}(1986){Braun}, {Goss}, {Caswell}, \&
  {Roger}}]{Braun86}
{Braun}, R., {Goss}, W.~M., {Caswell}, J.~L., \& {Roger}, R.~S. 1986, \aap,
  162, 259

\bibitem[{{Braun} {et~al.}(1987){Braun}, {Gull}, \& {Perley}}]{Braun87}
{Braun}, R., {Gull}, S.~F., \& {Perley}, R.~A. 1987,
  \href{http://dx.doi.org/10.1038/327395a0}{\nat, 327, 395}

\bibitem[{{Burrows} {et~al.}(2007){Burrows}, {Dessart}, {Ott}, \&
  {Livne}}]{Burrows07}
{Burrows}, A., {Dessart}, L., {Ott}, C.~D., \& {Livne}, E. 2007,
  \href{http://dx.doi.org/10.1016/j.physrep.2007.02.001}{\physrep, 442, 23}

\bibitem[Burrows \& Hayes(1996)]{Burrows96} Burrows, A., \& Hayes, J.\
  1996, Physical Review Letters, 76, 352

\bibitem[{{Burrows} {et~al.}(1995){Burrows}, {Hayes}, \& {Fryxell}}]{Burrows95}
{Burrows}, A., {Hayes}, J., \& {Fryxell}, B.~A. 1995,
  \href{http://dx.doi.org/10.1086/176188}{\apj, 450, 830}

\bibitem[{{Burrows} {et~al.}(2006){Burrows}, {Livne}, {Dessart}, {Ott}, \&
  {Murphy}}]{Burrows06}
{Burrows}, A., {Livne}, E., {Dessart}, L., {Ott}, C.~D., \& {Murphy}, J. 2006,
  \href{http://dx.doi.org/10.1086/500174}{\apj, 640, 878}

\bibitem[Burrows et al.(2005)]{Burrows05} Burrows, A., Walder, 
R., Ott, C.~D., 
\& Livne, E.\ 2005, The Fate of the Most Massive Stars, 332, 350

\bibitem[Chevalier(2005)]{Chevalier05} Chevalier, R.~A.\ 2005, 
\apj, 619, 839

\bibitem[{{Chevalier} \& {Kirshner}(1979)}]{Chevalier79}
{Chevalier}, R.~A., \& {Kirshner}, R.~P. 1979,
  \href{http://dx.doi.org/10.1086/157377}{\apj, 233, 154}

\bibitem[{{Chevalier} \& {Oishi}(2003)}]{Chevalier03Clumpy}
{Chevalier}, R.~A., \& {Oishi}, J. 2003,
  \href{http://dx.doi.org/10.1086/377572}{\apjl, 593, L23}

\bibitem[Claeys et al.(2011)]{Claeys11} Claeys, J.~S.~W., 
   de Mink, S.~E., Pols, O.~R., 
   Eldridge, J.~J., \& Baes, M.\ 2011, \aap, 528, A131 

\bibitem[{{Delaney}(2004)}]{DeLaney04}
{Delaney}, T.~A. 2004, PhD thesis, Univ.\ of Minnesota

\bibitem[{{DeLaney} {et~al.}(2010){DeLaney}, {Rudnick}, {Stage}, {Smith},
  {Isensee}, {Rho}, {Allen}, {Gomez}, {Kozasa}, {Reach}, {Davis}, \&
  {Houck}}]{DeLaney10}
{DeLaney}, T., {Rudnick}, L., {Stage}, M.~D., {et~al.} 2010,
  \href{http://dx.doi.org/10.1088/0004-637X/725/2/2038}{\apj, 725, 2038}

\bibitem[Ennis et al.(2006)]{Ennis06} Ennis, J.~A., Rudnick, 
  L., Reach, W.~T., et al.\ 2006, \apj, 652, 376 

\bibitem[Eriksen et al.(2009)]{Eriksen09} Eriksen, K.~A., Arnett, 
D., McCarthy, D.~W., \& Young, P.\ 2009, \apj, 697, 29 

\bibitem[Eriksen et al.(2001)]{Eriksen01} Eriksen, K.~A., Morse, 
J.~A., Kirshner, R.~P., 
\& Winkler, P.~F.\ 2001, Young Supernova Remnants, 565, 193 

\bibitem[{{Fesen}(2001)}]{Fesen01Turb}
{Fesen}, R.~A. 2001, \href{http://dx.doi.org/10.1086/319181}{\apjs, 133, 161}

\bibitem[{{Fesen} \& {Becker}(1991)}]{Fesen91}
{Fesen}, R.~A., \& {Becker}, R.~H. 1991,
  \href{http://dx.doi.org/10.1086/169926}{\apj, 371, 621}

\bibitem[{{Fesen} \& {Gunderson}(1996)}]{Fesen96}
{Fesen}, R.~A., \& {Gunderson}, K.~S. 1996,
  \href{http://dx.doi.org/10.1086/177923}{\apj, 470, 967}

\bibitem[{{Fesen} {et~al.}(2001){Fesen}, {Morse}, {Chevalier}, {Borkowski},
  {Gerardy}, {Lawrence}, \& {van den Bergh}}]{Fesen01WFPC2}
{Fesen}, R.~A., {Morse}, J.~A., {Chevalier}, R.~A., {et~al.} 2001,
  \href{http://dx.doi.org/10.1086/323539}{\aj, 122, 2644}

\bibitem[Fesen et al.(2011)]{2011ApJ...736..109F} Fesen, R.~A., Zastrow, 
J.~A., Hammell, M.~C., Shull, J.~M., \& Silvia, D.~W.\ 2011, \apj, 736, 109 

\bibitem[{{Fesen} {et~al.}(2006{\natexlab{a}}){Fesen}, {Hammell}, {Morse},
  {Chevalier}, {Borkowski}, {Dopita}, {Gerardy}, {Lawrence}, {Raymond}, \& {van
  den Bergh}}]{Fesen06a}
{Fesen}, R.~A., {Hammell}, M.~C., {Morse}, J., {et~al.} 2006{\natexlab{a}},
  \href{http://dx.doi.org/10.1086/498092}{\apj, 636, 859}

\bibitem[{{Fesen} {et~al.}(2006{\natexlab{b}}){Fesen}, {Hammell}, {Morse},
  {Chevalier}, {Borkowski}, {Dopita}, {Gerardy}, {Lawrence}, {Raymond}, \& {van
  den Bergh}}]{Fesen06bowtie}
{Fesen}, R.~A., {Hammell}, M.~C., {Morse}, J., {et~al.} 2006{\natexlab{b}},
  \href{http://dx.doi.org/10.1086/504254}{\apj, 645, 283}

\bibitem[{{Fryer} \& {Warren}(2004)}]{Fryer04}
{Fryer}, C.~L., \& {Warren}, M.~S. 2004,
  \href{http://dx.doi.org/10.1086/380193}{\apj, 601, 391}

\bibitem[{{Hammer} {et~al.}(2010){Hammer}, {Janka}, \& {M{\"u}ller}}]{Hammer10}
{Hammer}, N.~J., {Janka}, H., \& {M{\"u}ller}, E. 2010,
  \href{http://dx.doi.org/10.1088/0004-637X/714/2/1371}{\apj, 714, 1371}

\bibitem[Hanke et al.(2012)]{Hanke12} Hanke, F., Marek, A., 
          M{\"u}ller, B., \& Janka, H.-T.\ 2012, \apj, 755, 138 

\bibitem[{{Hines} {et~al.}(2004){Hines}, {Rieke}, {Gordon}, {Rho}, {Misselt},
  {Woodward}, {Werner}, {Krause}, {Latter}, {Engelbracht}, {Egami}, {Kelly},
  {Muzerolle}, {Stansberry}, {Su}, {Morrison}, {Young}, {Noriega-Crespo},
  {Padgett}, {Gehrz}, {Polomski}, {Beeman}, \& {Haller}}]{Hines04}
{Hines}, D.~C., {Rieke}, G.~H., {Gordon}, K.~D., {et~al.} 2004,
  \href{http://dx.doi.org/10.1086/422583}{\apjs, 154, 290}

\bibitem[Ho \& Heinke(2009)]{Ho09} Ho, W.~C.~G., \&
  Heinke, C.~O.\ 2009, \nat, 462, 71

\bibitem[Hughes et al.(2000)]{Hughes00} Hughes, J.~P., Rakowski, 
C.~E., Burrows, D.~N., \& Slane, P.~O.\ 2000, \apjl, 528, L109

\bibitem[{{Hwang} \& {Laming}(2012)}]{Hwang12}
{Hwang}, U., \& {Laming}, J.~M. 2012,
  \href{http://dx.doi.org/10.1088/0004-637X/746/2/130}{\apj, 746, 130}

\bibitem[Hwang \& Laming(2003)]{Hwang03} Hwang, U., \&
Laming, J.~M.\ 2003, \apj, 597, 362

\bibitem[{{Hwang} {et~al.}(2004){Hwang}, {Laming}, {Badenes}, {Berendse},
  {Blondin}, {Cioffi}, {DeLaney}, {Dewey}, {Fesen}, {Flanagan}, {Fryer},
  {Ghavamian}, {Hughes}, {Morse}, {Plucinsky}, {Petre}, {Pohl}, {Rudnick},
  {Sankrit}, {Slane}, {Smith}, {Vink}, \& {Warren}}]{Hwang04}
{Hwang}, U., {Laming}, J.~M., {Badenes}, C., {et~al.} 2004,
  \href{http://dx.doi.org/10.1086/426186}{\apjl, 615, L117}

\bibitem[{{Isensee} {et~al.}(2010){Isensee}, {Rudnick}, {DeLaney}, {Smith},
  {Rho}, {Reach}, {Kozasa}, \& {Gomez}}]{Isensee10}
{Isensee}, K., {Rudnick}, L., {DeLaney}, T., {et~al.} 2010,
  \href{http://dx.doi.org/10.1088/0004-637X/725/2/2059}{\apj, 725, 2059}

\bibitem[Janka(2012)]{Janka12} Janka, H.-T.\ 2012, Annual 
Review of Nuclear and Particle Science, 62, 407 

\bibitem[Janka et al.(2003)]{Janka03} Janka, H.-T., Buras, R., 
\& Rampp, M.\ 2003, Nuclear Physics A, 718, 269 

\bibitem[{{Janka} {et~al.}(2007){Janka}, {Langanke}, {Marek},
  {Mart{\'{\i}}nez-Pinedo}, \& {M{\"u}ller}}]{Janka07}
{Janka}, H., {Langanke}, K., {Marek}, A., {Mart{\'{\i}}nez-Pinedo}, G., \&
  {M{\"u}ller}, B. 2007,
  \href{http://dx.doi.org/10.1016/j.physrep.2007.02.002}{\physrep, 442, 38}

\bibitem[{{Kamper} \& {van den Bergh}(1976)}]{Kamper76}
{Kamper}, K., \& {van den Bergh}, S. 1976,
  \href{http://dx.doi.org/10.1086/190400}{\apjs, 32, 351}

\bibitem[{{Keohane} {et~al.}(1996){Keohane}, {Rudnick}, \&
  {Anderson}}]{Keohane96}
{Keohane}, J.~W., {Rudnick}, L., \& {Anderson}, M.~C. 1996,
  \href{http://dx.doi.org/10.1086/177511}{\apj, 466, 309}

\bibitem[{{Khokhlov} {et~al.}(1999){Khokhlov}, {H{\"o}flich}, {Oran},
  {Wheeler}, {Wang}, \& {Chtchelkanova}}]{Khokhlov99}
{Khokhlov}, A.~M., {H{\"o}flich}, P.~A., {Oran}, E.~S., {et~al.} 1999,
  \href{http://dx.doi.org/10.1086/312305}{\apjl, 524, L107}

\bibitem[{{Kifonidis} {et~al.}(2000){Kifonidis}, {Plewa}, {Janka}, \&
  {M{\"u}ller}}]{Kifonidis00}
{Kifonidis}, K., {Plewa}, T., {Janka}, H., \& {M{\"u}ller}, E. 2000,
  \href{http://dx.doi.org/10.1086/312541}{\apjl, 531, L123}

\bibitem[{{Kifonidis} {et~al.}(2003){Kifonidis}, {Plewa}, {Janka}, \&
  {M{\"u}ller}}]{Kifonidis03}
{Kifonidis}, K., {Plewa}, T., {Janka}, H., \& {M{\"u}ller}, E. 2003,
  \href{http://dx.doi.org/10.1051/0004-6361:20030863}{\aap, 408, 621}

\bibitem[{{Kj{\ae}r} {et~al.}(2010){Kj{\ae}r}, {Leibundgut}, {Fransson},
  {Jerkstrand}, \& {Spyromilio}}]{Kjaer10}
{Kj{\ae}r}, K., {Leibundgut}, B., {Fransson}, C., {Jerkstrand}, A., \&
  {Spyromilio}, J. 2010,
  \href{http://dx.doi.org/10.1051/0004-6361/201014538}{\aap, 517, A51}

\bibitem[{{Kotake} {et~al.}(2005){Kotake}, {Yamada}, \& {Sato}}]{Kotake05}
{Kotake}, K., {Yamada}, S., \& {Sato}, K. 2005,
  \href{http://dx.doi.org/10.1086/425911}{\apj, 618, 474}

\bibitem[{{Krause} {et~al.}(2008){Krause}, {Birkmann}, {Usuda}, {Hattori},
  {Goto}, {Rieke}, \& {Misselt}}]{Krause08}
{Krause}, O., {Birkmann}, S.~M., {Usuda}, T., {et~al.} 2008,
  \href{http://dx.doi.org/10.1126/science.1155788}{Science, 320, 1195}

\bibitem[Krause et al.(2008)]{Krause08b} Krause, O., Tanaka, M., 
Usuda, T., et al.\ 2008, \nat, 456, 617 

\bibitem[{{Laming} \& {Hwang}(2003)}]{Laming03}
{Laming}, J.~M., \& {Hwang}, U. 2003,
  \href{http://dx.doi.org/10.1086/378268}{\apj, 597, 347}

\bibitem[{{Laming} {et~al.}(2006){Laming}, {Hwang}, {Radics}, {Lekli}, \&
  {Tak{\'a}cs}}]{Laming06}
{Laming}, J.~M., {Hwang}, U., {Radics}, B., {Lekli}, G., \& {Tak{\'a}cs}, E.
  2006, \href{http://dx.doi.org/10.1086/503553}{\apj, 644, 260}

\bibitem[{{Lawrence} {et~al.}(1995){Lawrence}, {MacAlpine}, {Uomoto},
  {Woodgate}, {Brown}, {Oliversen}, {Lowenthal}, \& {Liu}}]{Lawrence95}
{Lawrence}, S.~S., {MacAlpine}, G.~M., {Uomoto}, A., {et~al.} 1995,
  \href{http://dx.doi.org/10.1086/117477}{\aj, 109, 2635}

\bibitem[{{Li} {et~al.}(1993){Li}, {McCray}, \& {Sunyaev}}]{Li93}
{Li}, H., {McCray}, R., \& {Sunyaev}, R.~A. 1993,
  \href{http://dx.doi.org/10.1086/173534}{\apj, 419, 824}

\bibitem[{{Maeda} {et~al.}(2008){Maeda}, {Kawabata}, {Mazzali}, {Tanaka},
  {Valenti}, {Nomoto}, {Hattori}, {Deng}, {Pian}, {Taubenberger}, {Iye},
  {Matheson}, {Filippenko}, {Aoki}, {Kosugi}, {Ohyama}, {Sasaki}, \&
  {Takata}}]{Maeda08}
{Maeda}, K., {Kawabata}, K., {Mazzali}, P.~A., {et~al.} 2008,
  \href{http://dx.doi.org/10.1126/science.1149437}{Science, 319, 1220}

\bibitem[Marek \& Janka(2009)]{MK09} Marek, A., \& Janka, H.-T.\ 2009, \apj, 694, 664 

\bibitem[{{Markert} {et~al.}(1983){Markert}, {Clark}, {Winkler}, \&
  {Canizares}}]{Markert83}
{Markert}, T.~H., {Clark}, G.~W., {Winkler}, P.~F., \& {Canizares}, C.~R. 1983,
  \href{http://dx.doi.org/10.1086/160939}{\apj, 268, 134}

\bibitem[Masada et al.(2012)]{Masada12} Masada, Y., Takiwaki, 
T., Kotake, K., \& Sano, T.\ 2012, \apj, 759, 110

\bibitem[{{Massey} \& {Gronwall}(1990)}]{Massey90}
{Massey}, P., \& {Gronwall}, C. 1990,
  \href{http://dx.doi.org/10.1086/168991}{\apj, 358, 344}

\bibitem[{{Mazzali} {et~al.}(2005){Mazzali}, {Kawabata}, {Maeda}, {Nomoto},
  {Filippenko}, {Ramirez-Ruiz}, {Benetti}, {Pian}, {Deng}, {Tominaga},
  {Ohyama}, {Iye}, {Foley}, {Matheson}, {Wang}, \& {Gal-Yam}}]{Mazzali05}
{Mazzali}, P.~A., {Kawabata}, K.~S., {Maeda}, K., {et~al.} 2005,
  \href{http://dx.doi.org/10.1126/science.1111384}{Science, 308, 1284}

\bibitem[Mazzali et al.(2007)]{Mazzali07} Mazzali, P.~A., 
Kawabata, K.~S., Maeda, K., et al.\ 2007, \apj, 670, 592 

\bibitem[Milisavljevic et al.(2012)]{Milisavljevic12} Milisavljevic, 
D., Fesen, R.~A., Chevalier, R.~A., et al.\ 2012, \apj, 751, 25 

\bibitem[Milisavljevic et al.(2010)]{Milisavljevic10}
  Milisavljevic, D., Fesen, R.~A., Gerardy, C.~L., Kirshner, R.~P., \&
  Challis, P.\ 2010, \apj, 709, 1343

\bibitem[Minkowski(1959)]{Minkowski59} Minkowski, R.\ 1959, URSI 
Symp.~1: Paris Symposium on Radio Astronomy, 9, 315 

\bibitem[{{Minkowski}(1968)}]{Minkowski68}
{Minkowski}, R. 1968, {Nonthermal Galactic Radio Sources}, ed. {Middlehurst,
  B.~M.~\& Aller, L.~H.} (the University of Chicago Press), 623

\bibitem[{{Modjaz} {et~al.}(2008){Modjaz}, {Kirshner}, {Blondin}, {Challis}, \&
  {Matheson}}]{Modjaz08}
{Modjaz}, M., {Kirshner}, R.~P., {Blondin}, S., {Challis}, P., \& {Matheson},
  T. 2008, \href{http://dx.doi.org/10.1086/593135}{\apjl, 687, L9}

\bibitem[{{Nordhaus} {et~al.}(2010){Nordhaus}, {Burrows}, {Almgren}, \&
  {Bell}}]{Nordhaus10}
{Nordhaus}, J., {Burrows}, A., {Almgren}, A., \& {Bell}, J. 2010,
  \href{http://dx.doi.org/10.1088/0004-637X/720/1/694}{\apj, 720, 694}

\bibitem[{{Park} {et~al.}(2002){Park}, {Roming}, {Hughes}, {Slane}, {Burrows},
  {Garmire}, \& {Nousek}}]{Park02}
{Park}, S., {Roming}, P.~W.~A., {Hughes}, J.~P., {et~al.} 2002,
  \href{http://dx.doi.org/10.1086/338861}{\apjl, 564, L39}

\bibitem[{{Reed} {et~al.}(1995){Reed}, {Hester}, {Fabian}, \&
  {Winkler}}]{Reed95}
{Reed}, J.~E., {Hester}, J.~J., {Fabian}, A.~C., \& {Winkler}, P.~F. 1995,
  \href{http://dx.doi.org/10.1086/175308}{\apj, 440, 706}

\bibitem[Rest et al.(2011)]{Rest11} Rest, A., Foley, R.~J., 
Sinnott, B., et al.\ 2011, \apj, 732, 3 

\bibitem[Rest et al.(2008)]{Rest08} Rest, A., Matheson, T., 
Blondin, S., et al.\ 2008, \apj, 680, 1137 

\bibitem[Rest et al.(2005)]{Rest05} Rest, A., Suntzeff, N.~B., 
Olsen, K., et al.\ 2005, \nat, 438, 1132 

\bibitem[Reynoso et al.(1997)]{Reynoso97} Reynoso, E.~M., 
 Goss, W.~M., Dubner, G.~M., Winkler, P.~F., \& Schwarz, U.~J.\ 1997, \aap, 317, 203 

\bibitem[{{Scheck} {et~al.}(2006){Scheck}, {Kifonidis}, {Janka}, \&
  {M{\"u}ller}}]{Scheck06}
{Scheck}, L., {Kifonidis}, K., {Janka}, H., \& {M{\"u}ller}, E. 2006,
  \href{http://dx.doi.org/10.1051/0004-6361:20064855}{\aap, 457, 963}

\bibitem[Shibata et al.(2006)]{Shibata06} Shibata, M., Liu, 
Y.~T., Shapiro, S.~L., \& Stephens, B.~C.\ 2006, \prd, 74, 104026

\bibitem[{{Smith} {et~al.}(2009){Smith}, {Rudnick}, {Delaney}, {Rho}, {Gomez},
  {Kozasa}, {Reach}, \& {Isensee}}]{Smith09}
{Smith}, J.~D.~T., {Rudnick}, L., {Delaney}, T., {et~al.} 2009,
  \href{http://dx.doi.org/10.1088/0004-637X/693/1/713}{\apj, 693, 713}

\bibitem[{{Stone}(1977)}]{Stone77}
{Stone}, R.~P.~S. 1977, \href{http://dx.doi.org/10.1086/155732}{\apj, 218, 767}

\bibitem[Tanaka et al.(2012)]{Tanaka12} Tanaka, M., Kawabata, 
K.~S., Hattori, T., et al.\ 2012, \apj, 754, 63 

\bibitem[{{Tananbaum}(1999)}]{Tananbaum99}
{Tananbaum}, H. 1999, \iaucirc, 7246, 1

\bibitem[Taubenberger et al.(2009)]{Taubenberger09} Taubenberger, S., 
Valenti, S., Benetti, S., et al.\ 2009, \mnras, 397, 677 

\bibitem[{{Thorstensen} {et~al.}(2001){Thorstensen}, {Fesen}, \& {van den
  Bergh}}]{Thorstensen01}
{Thorstensen}, J.~R., {Fesen}, R.~A., \& {van den Bergh}, S. 2001,
  \href{http://dx.doi.org/10.1086/321138}{\aj, 122, 297}

\bibitem[Ugliano et al.(2012)]{Ugliano12} Ugliano, M., Janka, 
         H.-T., Marek, A., \& Arcones, A.\ 2012, \apj, 757, 69 

\bibitem[{{van den Bergh} \& {Dodd}(1970)}]{vandenBergh70}
{van den Bergh}, S., \& {Dodd}, W.~W. 1970,
  \href{http://dx.doi.org/10.1086/150681}{\apj, 162, 485}

\bibitem[{{van den Bergh} \& {Kamper}(1983)}]{vandenBergh83}
{van den Bergh}, S., \& {Kamper}, K.~W. 1983,
  \href{http://dx.doi.org/10.1086/160938}{\apj, 268, 129}

\bibitem[van den Bergh \& Kamper(1985)]{BK85} van den Bergh, S., \& Kamper, K.\ 1985, \apj, 293, 537 

\bibitem[{{Vogt} \& {Dopita}(2010)}]{Vogt10}
{Vogt}, F., \& {Dopita}, M.~A. 2010,
  \href{http://dx.doi.org/10.1088/0004-637X/721/1/597}{\apj, 721, 597}

\bibitem[{{Vogt} \& {Dopita}(2011)}]{Vogt11}
{Vogt}, F., \& {Dopita}, M.~A. 2011,
  \href{http://dx.doi.org/10.1007/s10509-010-0479-7}{\apss, 331, 521}

\bibitem[{{Wang} \& {Wheeler}(2008)}]{Wang08}
{Wang}, L., \& {Wheeler}, J.~C. 2008,
  \href{http://dx.doi.org/10.1146/annurev.astro.46.060407.145139}{\araa, 46,
  433}

\bibitem[{{Wang} {et~al.}(2002){Wang}, {Wheeler}, {H{\"o}flich}, {Khokhlov},
  {Baade}, {Branch}, {Challis}, {Filippenko}, {Fransson}, {Garnavich},
  {Kirshner}, {Lundqvist}, {McCray}, {Panagia}, {Pun}, {Phillips}, {Sonneborn},
  \& {Suntzeff}}]{Wang02}
{Wang}, L., {Wheeler}, J.~C., {H{\"o}flich}, P., {et~al.} 2002,
  \href{http://dx.doi.org/10.1086/342824}{\apj, 579, 671}

\bibitem[Wheeler et al.(2008)]{Wheeler08} Wheeler, J.~C., Maund, 
J.~R., \& Couch, S.~M.\ 2008, \apj, 677, 1091 

\bibitem[{{Wheeler} {et~al.}(2002){Wheeler}, {Meier}, \& {Wilson}}]{Wheeler02}
{Wheeler}, J.~C., {Meier}, D.~L., \& {Wilson}, J.~R. 2002,
  \href{http://dx.doi.org/10.1086/338953}{\apj, 568, 807}

\bibitem[{{Willingale} {et~al.}(2002){Willingale}, {Bleeker}, {van der Heyden},
  {Kaastra}, \& {Vink}}]{Willingale02}
{Willingale}, R., {Bleeker}, J.~A.~M., {van der Heyden}, K.~J., {Kaastra},
  J.~S., \& {Vink}, J. 2002,
  \href{http://dx.doi.org/10.1051/0004-6361:20011614}{\aap, 381, 1039}

\bibitem[{{Winkler} {et~al.}(2009){Winkler}, {Twelker}, {Reith}, \&
  {Long}}]{Winkler09}
{Winkler}, P.~F., {Twelker}, K., {Reith}, C.~N., \& {Long}, K.~S. 2009,
  \href{http://dx.doi.org/10.1088/0004-637X/692/2/1489}{\apj, 692, 1489}

\bibitem[{{Woosley} {et~al.}(1993){Woosley}, {Langer}, \& {Weaver}}]{Woosley93}
{Woosley}, S.~E., {Langer}, N., \& {Weaver}, T.~A. 1993,
  \href{http://dx.doi.org/10.1086/172886}{\apj, 411, 823}

\bibitem[{{Woosley} \& {Janka}(2005)}]{Woosley05}
{Woosley}, S., \& {Janka}, T. 2005,
  \href{http://dx.doi.org/10.1038/nphys172}{Nature Physics, 1, 147}

\bibitem[{{Young} {et~al.}(2006){Young}, {Fryer}, {Hungerford}, {Arnett},
  {Rockefeller}, {Timmes}, {Voit}, {Meakin}, \& {Eriksen}}]{Young06}
{Young}, P.~A., {Fryer}, C.~L., {Hungerford}, A., {et~al.} 2006,
  \href{http://dx.doi.org/10.1086/500108}{\apj, 640, 891}

\end{thebibliography}
\end{document}